\documentclass[11pt,a4paper]{article}
\usepackage[utf8]{inputenc}
\usepackage{amsmath}
\usepackage{placeins}
\usepackage{url}
\usepackage{natbib}
\usepackage{color,xcolor,setspace,geometry}
\setcitestyle{authoryear,open={(},close={)}}
\usepackage{cite}
\usepackage{amsfonts}
\usepackage[normalem]{ulem}
\usepackage{pifont}
\usepackage{hyperref}

\usepackage{arydshln}
\usepackage{multirow}

\usepackage{tikz}
\usetikzlibrary{shapes}
\usetikzlibrary{plotmarks}
\usetikzlibrary{tikzmark,matrix,calc}
\usetikzlibrary{arrows.meta,calc,decorations.markings,math,arrows.meta}

\usepackage{amssymb}
\usepackage{multirow}
\usepackage{graphicx}
\usepackage{soul}
\soulregister\citep7
\soulregister\cite7
\soulregister\citet7
\soulregister\citealp7
\soulregister\ref7
\usepackage{rotating}
\usepackage{setspace}
\usepackage{threeparttable}
\usepackage{authblk}
\usepackage{subfigure}

\usepackage{appendix}
\usepackage{bm}

\begin{document}
\begin{spacing}{1.5}	

\title{Predicting Qualification Thresholds in the UEFA Champions League and Europa League under the New League Phase Format}

\author[$\dag$ $\star$]{David Winkelmann}
\author[$\S$]{Rouven Michels}
\author[$\ddag$]{Christian Deutscher}

\affil[$\dag$]{\small Department of Business Decisions and Analytics\\ University of Vienna, Vienna, Austria}
\affil[$\S$]{ Department of Statistics\\ TU Dortmund, Dortmund, Germany}
\affil[$\ddag$]{ Department of Psychology and Sports Science\\ Bielefeld University, Bielefeld, Germany}

\affil[$\star$]{Corresponding author: david.winkelmann@uni-bielefeld.de}

\maketitle
\noindent
\normalsize

\begin{abstract}
For the 2024/25 season, the Union of European Football Associations (UEFA) introduced an incomplete round-robin format in the Champions League and Europa League, replacing the traditional group stage with a single league table of all 36 teams. Under this structure, the top eight teams advance directly to the Round of 16, while teams ranked 9th–24th qualify for a play-off round. Simulation-based analyses, such as those by commercial data analyst \textit{Opta}, provided indicative point thresholds for qualification but reveal deviations when compared with actual outcomes in the first season. To address these discrepancies, we employ a bivariate Dixon--Coles model that accounts for the lower frequency of draws observed in the 2024/25 and 2025/26 seasons, particularly in the Champions League, potentially driven by reduced incentives for teams to play for a draw. We proxy team strengths by Elo ratings and fit the model to different settings. This enables us to simulate match outcomes and to estimate qualification thresholds for both direct advancement and play-off participation. Our results provide scientific guidance for clubs and managers, supporting strategic decision-making under uncertainty regarding their progression prospects in the new format of UEFA club competitions.

\textbf{Keywords}:
Bivariate Poisson model, Dixon--Coles model, Elo rating, Incomplete round-robin tournament, Sports forecasting
\end{abstract}

\section{Introduction}
\label{sec:introductio}
The UEFA Champions League (UCL) is the most prestigious competition in European club football, drawing global audiences, generating substantial commercial investment, and showcasing the highest level of competitive play. Structural changes to its format, therefore, carry far-reaching implications -- not only for sporting fairness and competitive balance but also for the strategic planning of participating clubs, broadcasters, and sponsors. The 2024/25 season marked one of the most significant reforms in the tournament’s history: UEFA replaced its long-standing group stage format with the so-called \textit{incomplete round-robin} format. Now, 36 clubs compete in a single league phase, fundamentally reshaping the competition dynamics. Notably, the incomplete round-robin format has also been introduced in the UEFA Europa League (UEL) and UEFA Conference League, broadening its overall impact. This paper focuses on the UCL and UEL; the UEFA Conference League is not examined in detail due to its slightly different structure, with only six league-phase matches instead of eight per club.

Under the new system, all clubs within a competition are ranked in a single table. The top eight teams qualify directly for the Round of 16, while those ranked 9th to 24th enter play-offs for the remaining spots. Teams finishing 25th or lower are eliminated. Progression, therefore, requires a top-eight finish or success in the play-offs, thereby increasing the competitive premium of finishing near the top compared with the previous format, in which the second place in a group merely implied a potentially stronger opponent in the Round of 16 rather than the need to play an additional play-off round. At the same time, the new structure creates additional opportunities for lower-ranked teams, as even a 24th-place finish results in knockout stage participation. According to \citet{gyimesi2024competitive}, the format alteration may enhance competitive balance and yield fewer stakeless matches, though more (potentially) collusive matches may occur in the last round \citep{csato2026probabilistic,devriesere2026groups}. 

This change in tournament design potentially also strengthens the incentive to pursue victories in single matches \citep{csato2025hidden}. Under the old group stage format, a draw meant that one of three direct opponents collected only one point. In contrast, in the league-phase system, teams compete against 35 others in the same table, which diminishes the relative value of a draw. Similar to the introduction of awarding three points for a win instead of two \citep{dilger2009three}, the format change intends to enhance competitiveness and spectator appeal. While previous literature reports mixed evidence on the effectiveness of the three-point rule \citep{guedes2002changing}, early observations from the 2024/25 UCL season suggest stronger incentives for teams to seek a win, as we observe a considerably lower number of draws (see the descriptive statistics in Section~\ref{sec:data}).

Given the strategic implications of the new format and its altered ranking mechanism, understanding qualification thresholds required for direct or play-off entry in the incomplete round-robin format is of crucial importance. In particular, the frequency of draws plays a decisive role in shaping these thresholds, as highlighted in prior studies on competitive balance in sports \citep{fry2021managing, ficcadenti2023rank}. Moreover, since the final two league-phase matches are scheduled after the winter break, clubs must make transfer-market decisions under uncertainty regarding their progression prospects. This further underscores the need for reliable estimates of qualification thresholds in the UCL.

\begin{figure}[ht]
    \centering
    \includegraphics[width=0.5\linewidth]{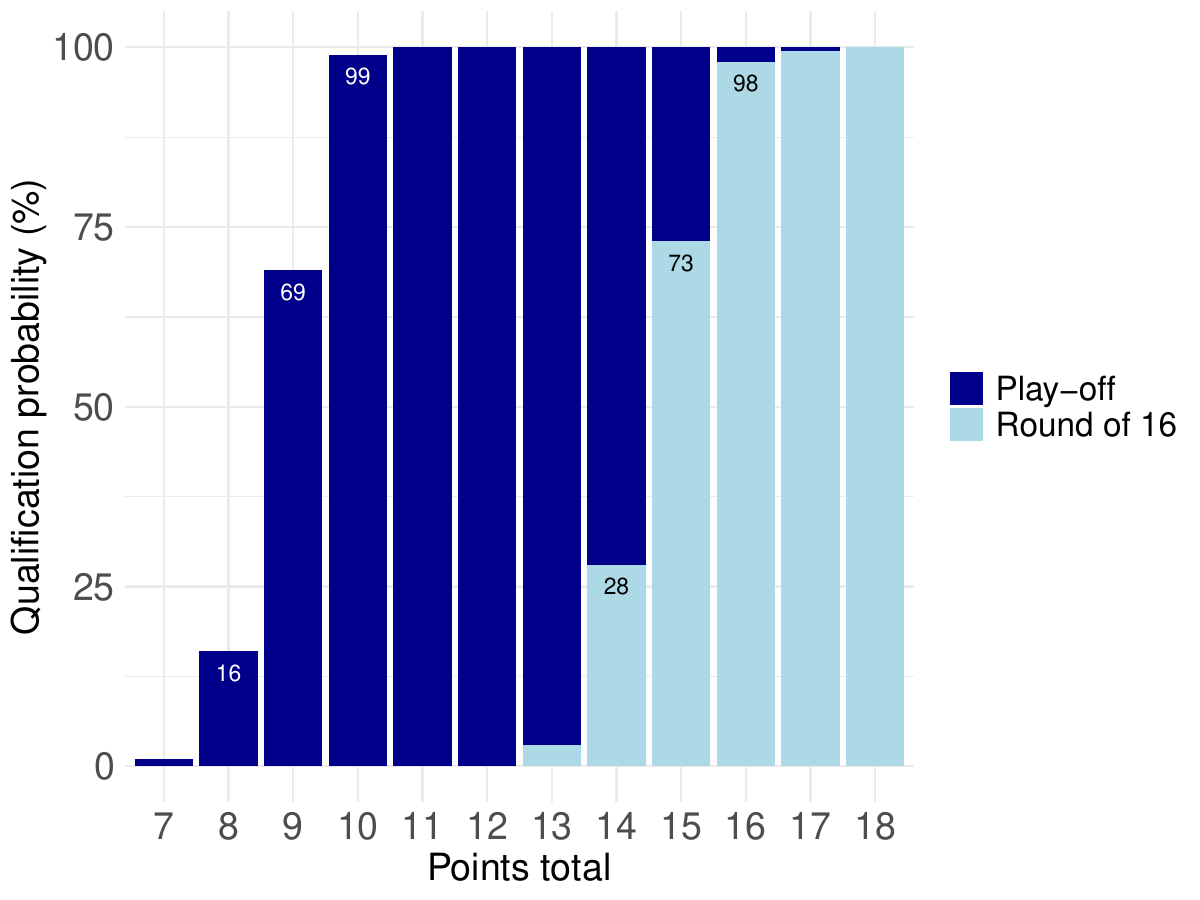}
    \caption{Likelihood of progressing to the play-off round and directly to the Round of 16, respectively, depending on the number of points obtained in the league phase. Own illustration based on \citet{opta}.}
    \label{fig:opta}
\end{figure}

The simulation-based prediction of qualification probabilities is well established in the literature (see, e.g.\ \citealp{groll2015prediction}, for the FIFA World Cup group stage, and \citealp{karlis2011robust}, for the UCL under the previous format). While there are some studies on UEFA's new competition structures (see, e.g.\ \citealp{csato2025does,devriesere2026groups,gyimesi2024competitive}), none of them accounts for the lower frequency of draws in the new format. Instead, insights on qualification thresholds have been provided primarily by commercial analytics firms to date. For instance, \textit{Opta} employed its proprietary simulation framework -- referred to as its ``supercomputer'' -- to estimate the points required for progression \citep{opta}. Their forecast of the 2024/25 Champions League season indicated that ``15 points is likely to be enough to progress to the last 16'' and that ``10 points will as good as guarantee a place in the top 24 places''. More specifically, 15 points were associated with a 73\% probability of securing a top-eight finish. In comparison, 10 points ensured qualification for the play-off round with a 99\% likelihood (see Figure~\ref{fig:opta}).

\begin{table}[ht]
    \centering
    \scalebox{0.7}{
    \begin{tabular}{cc|c|c|ccc|ccc|c}
         & & & & & & & Goals & Goals & Goal & \\
         & Place & Team & Matches & Wins & Draws & Losses & scored & conceded & difference & Points \\
         \hline
         \multirow{5}{*}{\rotatebox{90}{\parbox{2.5cm}{\centering Round of 16}}}
         & 1 & Liverpool FC & 8 & 7 & 0 & 1 & 17 & 5 & 12 & 21 \\
         & 2 & FC Barcelona & 8 & 6 & 1 & 1 & 28 & 13 & 15 & 19 \\
         & $\cdots$ & $\cdots$ & $\cdots$ & & $\cdots$ & & & $\cdots$ & & $\cdots$ \\
         & 7 & LOSC Lille & 8 & 5 & 1 & 2 & 17 & 10 & 7 & 16 \\
         & 8 & Aston Villa & 8 & 5 & 1 & 2 & 13 & 6 & 7 & 16 \\
         \hline
         \multirow{5}{*}{\rotatebox{90}{\parbox{2.5cm}{\centering Play-offs}}}
         & 9 & Atalanta BC & 8 & 4 & 3 & 1 & 20 & 6 & 14 & 15 \\
         & 10 & Borussia Dortmund & 8 & 5 & 0 & 3 & 22 & 12 & 10 & 15 \\
         & $\cdots$ & $\cdots$ & $\cdots$ & & $\cdots$ & & & $\cdots$ & & $\cdots$ \\
         & 23 & Sporting CP & 8 & 3 & 2 & 3 & 13 & 12 & 1 & 11 \\
         & 24 & Club Brugge KV & 8 & 3 & 2 & 3 & 7 & 11 & --4 & 11 \\
         \hline
         \multirow{5}{*}{\rotatebox{90}{\parbox{2.5cm}{\centering Elimination}}}
         & 25 & GNK Dinamo Zagreb & 8 & 3 & 2 & 3 & 12 & 19 & --7 & 11 \\
         & 26 & VfB Stuttgart & 8 & 3 & 1 & 4 & 13 & 17 & --4 & 10 \\
         & $\cdots$ & $\cdots$ & $\cdots$ & & $\cdots$ & & & $\cdots$ & & $\cdots$ \\
         & 35 & Slovan Bratislava & 8 & 0 & 0 & 8 & 7 & 27 & --20 & 0 \\
         & 36 & BSC Young Boys & 8 & 0 & 0 & 8 & 3 & 24 & --21 & 0 \\
    \end{tabular}}
    \caption{Final standings after the UCL league phase 2024/25.}
    \label{tab:standings}
\end{table}

When comparing these pre-season predictions with the actual outcomes of the first seasons (see Table~\ref{tab:standings} for an extract of the final UCL standings following the 2024/25 league phase; final standings for the 2025/26 UCL league phase (Table~\ref{tab:standings_CL_2526}) and both UEL seasons (Table~\ref{tab:standings_EL}) are provided in Appendix~\ref{sec:app_UCL} and~\ref{sec:app_UEL}), several unexpected results are revealed. Despite the forecasted 73\% probability, teams in the UCL that collected 15 points in the 2024/25 season did not qualify directly for the Round of 16 but were instead required to compete in the play-off phase (in the UEL, one team with 14 points advanced directly to the Round of 16). Even more striking were the cases of GNK Dinamo Zagreb and VfB Stuttgart, ranked 25th and 26th. Pre-season predictions had suggested a 99\% probability (or higher) that the 10 and 11 points they achieved, respectively, would suffice to remain in the tournament. Yet, both were eliminated (in the UEL, two out of five teams with 10 points were eliminated). Reflecting on these discrepancies between forecasted and actual qualification thresholds, in a post-match interview with sports broadcaster DAZN, VfB Stuttgart’s manager Sebastian Hoeneß critically remarked that ``[\ldots] AI is probably not as great as everyone says'' when asked about Opta's pre-competition forecasts.

These discrepancies in the first UCL season under the new format highlight the need for a rigorous scientific framework to estimate qualification thresholds. This paper precisely simulates league outcomes while accounting for changes in incentive structures to determine qualification thresholds, thereby offering statistically grounded guidance to teams in the UCL and UEL. We apply the bivariate Dixon--Coles model introduced by \citet{dixon1997modelling}, which allows for capturing shifts in the probability distribution of scorelines and accommodates deviations in draw frequencies from their expected levels \citep{michels2025extending}. Adjusting for such deviations is crucial, as the frequency of draws can materially affect predicted qualification thresholds. To account for heterogeneous team strengths, we proxy unobserved abilities using Elo ratings (see, e.g.\ \citealp{bosker2024impact,hvattum2010using,yildirim2025new}). The predictive validity of Elo ratings in the UCL has been demonstrated by \citet{csato2024club}. We show that our model can better predict unseen UCL and UEL matches than competing models, forecasting the reduced number of draws more accurately and, as a consequence, providing reasonable qualification thresholds.

The remainder of the paper is structured as follows. Section~\ref{sec:literature} describes the UEFA club competitions and presents related literature. Section~\ref{sec:data} introduces the data, Section~\ref{sec:prediction} the modelling approach employed to simulate the new formats, while Section~\ref{sec:equal} presents the predicted qualification thresholds under different scenarios. Finally, Section~\ref{sec:conclusion} discusses the findings and their managerial implications.

\section{UEFA club competitions and related literature}
\label{sec:literature}
Statistical modelling and forecasting of football match outcomes is a well-established research area (\citealp{ley2019ranking, wunderlich2021forecasting}). The independent Poisson model (\citealp{maher1982modelling}) serves as the foundation for many modern approaches, while \citet{dixon1997modelling} introduced a dependence correction to better capture low-scoring outcomes. \citet{karlis2005bivariate} proposed a bivariate Poisson model to explicitly account for dependence between teams' goal counts, and \citet{michels2025extending} further extended the Dixon--Coles framework. These models form the methodological basis for estimating probabilities of football match outcomes. Their primary purpose is to enhance the enjoyment of fans by providing accurate predictions on match outcomes, while also serving as a basis for sports managers to evaluate performance or traders' decisions in the betting market \citep{corona2019bayesian}. For the UCL in particular, the literature has addressed a range of topics, including tournament designs, the effects of qualification systems, seeding regimes, and competitive balance. 

UEFA’s premier club competition, the European Champions Clubs’ Cup (European Cup), was inaugurated in 1955. Since then, its structure has been repeatedly adjusted to the evolving demands of national leagues, broadcasters, and sponsors, often to preserve a high degree of outcome uncertainty \citep{scarf2009numerical}. In 1992/93, the newly established UCL replaced the European Cup as UEFA’s flagship tournament. The overall tournament format remained unchanged for more than two decades preceding the 2024/25 season: the group stage consisted of eight groups of four teams, each playing a double round-robin \citep{rasmussen2008round}, with the top two advancing to the Round of 16. However, there were different minor changes in the design that were addressed in the literature.

In 2015, the UEFA implemented a change in the UCL seeding system \citep{csato2020uefa}. Previously, with the exception of the reigning UCL titleholder, who was always allocated to pot 1, placement in the seeding pots was determined by the club coefficient. From 2015 onwards, however, pot 1 contained the winners of the top seven domestic leagues, together with the winner of the previous UCL season \citep{dagaev2019seeding}. \citet{corona2019bayesian} show that this change increased uncertainty over progression from the group stage. From the 2018/19 season onwards, there was also a change in the UCL qualifying system. Specifically, the top four leagues received an additional fixed place in the UCL, while the number of places awarded to clubs participating in the qualification was reduced from ten to six. \citet{csato2022uefa} demonstrates that this change resulted in a considerable decrease in the probability of participation in the UCL for the winners of domestic leagues in low-ranked associations, with expected prize money decreasing by more than one million Euros for some clubs.

Studies by \citet{triguero2023competitive} and \citet{ramchandani2023you} report a decline in the ex-ante competitive balance of the UCL group stage in recent decades. By contrast, \citet{csato2025competitive} argue that these findings may be sensitive to the chosen measure and propose alternative metrics that challenge the earlier conclusions. Still, enhancing competitive balance was a central motivation for UEFA when introducing the new tournament format with a single league table. While the previous group stage format was structurally simple, it frequently produced non-competitive fixtures, including so-called dead rubbers, in the final rounds, as some teams had already secured qualification \citep{csato2024club}; for studies on the competitiveness of the last round matches in the UCL group stage, see, \citet{csato2024tournament,csato2026probabilistic, gyimesi2024competitive,devriesere2026groups}.

Unsurprisingly, the literature has devoted considerable attention to the fundamental change in the format introduced in the 2024/25 season. \citet{gyimesi2024competitive} shows an increase in competitive balance for the UCL, resulting in a higher degree of match uncertainty and fewer stakeless fixtures. Similarly, \citet{devriesere2026groups} and \citet{csato2026probabilistic} find that the incomplete round-robin format generates more competitive matches than the former group stage format, while \citet{guyon2025drawing} examine the implications of the scheduling procedure using simulation data. Moreover, \citet{csato2025does} show that the qualification prospects in both formats depend on the draw procedure as well as the degree of competitive balance among participants. Beyond these findings, \citet{csato2024club} demonstrates that replacing the current seeding regime, based on the UEFA club coefficient, with a variant of the Elo method could reduce the risk of unbalanced schedules. Nevertheless, \citet{csato2025ranking} argue that the newly introduced league phase format does not necessarily produce the most accurate ranking of clubs. A broader overview of tournament designs is provided by \citet{devriesere2025tournament}.

\section{Data}
\label{sec:data}
In this section, we provide descriptive statistics for the teams participating in the 2024/25 and 2025/26 UCL and UEL seasons, covering their strength via market values and Elo ratings. We then present summary statistics on match outcomes, comparing them with data from the last four UCL seasons under the old format, as well as with figures from the 2023/24 season of the four top European domestic leagues.

\subsection{Descriptive statistics for 2024/25 and 2025/26 UCL and UEL teams}
\label{sec:descriptive}
For the draw procedure, clubs are allocated into four seeding pots of nine teams each, determined by their UEFA club coefficient (see \citealp{guyon2025drawing} for an analysis of draw procedures in the incomplete round-robin format of the UCL, and \citealp{csato2025does} for a discussion of its implications). Each team plays against eight different opponents, two from each pot, with one match at home and one away. While in both UCL seasons Pot 1 consists exclusively of clubs from the top five national leagues, the remaining pots are more heterogeneous in composition. A similar structure applies to the UEL, where 36 clubs from 22 (2024/25) and 23 (2025/26) associations participate, respectively.

\begin{table}[h]
    \centering
    \scalebox{0.72}{
    \begin{tabular}{ccc|cccc|c|cccc|c}
         & & & \multicolumn{5}{c|}{Champions League} & \multicolumn{5}{c}{Europa League} \\
         Season & Variable & Statistic & Pot 1 & Pot 2 & Pot 3 & Pot 4 & All & Pot 1 & Pot 2 & Pot 3 & Pot 4 & All \\
        \hline

        \multirow{10}{*}{2024/25}
        & & Min & 462 & 137 & 63 & 21 & 21 & 78 & 22 & 19 & 10 & 10 \\
        & & Max & 1340 & 1170 & 393 & 595 & 1340 & 857 & 431 & 189 & 319 & 857 \\
        & \multirow{1}{*}{\shortstack{Market values\\(in Mio €)}} & Mean & 880 & 509 & 192 & 224 & 451 & 348 & 154 & 65 & 121 & 172 \\
        & & Median & 883 & 530 & 180 & 208 & 341 & 248 & 94 & 50 & 114 & 91 \\
        & & SD & 295 & 309 & 121 & 181 & 362 & 278 & 131 & 50 & 103 & 190 \\
        \cline{2-13}
        & & Min & 1861 & 1572 & 1553 & 1446 & 1446 & 1593 & 1475 & 1506 & 1245 & 1245 \\
        & & Max & 2060 & 1950 & 1835 & 1792 & 2060 & 1797 & 1746 & 1699 & 1752 & 1797 \\
        & \multirow{1}{*}{Elo ratings} & Mean & 1927 & 1807 & 1688 & 1708 & 1782 & 1731 & 1640 & 1594 & 1564 & 1632 \\
        & & Median & 1905 & 1836 & 1686 & 1769 & 1791 & 1774 & 1662 & 1599 & 1615 & 1636 \\
        & & SD & 65 & 115 & 106 & 115 & 137 & 79 & 81 & 62 & 159 & 117 \\

        \hline
        \hline

        \multirow{10}{*}{2025/26}
        & & Min & 438 & 173 & 54 & 13 & 13 & 55 & 28 & 48 & 26 & 26 \\
        & & Max & 1400 & 1370 & 890 & 747 & 1400 & 491 & 299 & 540 & 285 & 540 \\
        & \multirow{1}{*}{\shortstack{Market values\\(in Mio €)}} & Mean & 1011 & 501 & 320 & 220 & 513 & 232 & 115 & 139 & 122 & 153 \\
        & & Median & 1120 & 439 & 253 & 86 & 403 & 216 & 92 & 68 & 101 & 109 \\
        & & SD & 297 & 354 & 260 & 245 & 417 & 147 & 86 & 158 & 101 & 130 \\
        \cline{2-13}
        & & Min & 1820 & 1754 & 1655 & 1302 & 1302 & 1525 & 1507 & 1464 & 1487 & 1464 \\
        & & Max & 2010 & 1996 & 1846 & 1862 & 2010 & 1846 & 1761 & 1795 & 1751 & 1846 \\
        & \multirow{1}{*}{Elo ratings} & Mean & 1934 & 1827 & 1739 & 1654 & 1788 & 1703 & 1616 & 1584 & 1608 & 1628 \\
        & & Median & 1939 & 1821 & 1743 & 1711 & 1800 & 1735 & 1589 & 1540 & 1611 & 1620 \\
        & & SD & 52 & 70 & 69 & 175 & 145 & 120 & 79 & 112 & 93 & 108 \\

        \hline
    \end{tabular}}
    \caption{Summary statistics of market values (in million €) and Elo ratings for teams participating in the UEFA Champions League (UCL) and UEFA Europa League (UEL) as of 1 September 2024 (2024/25 season) and 1 September 2025 (2025/26 season), stratified by seeding pot.}
    \label{tab:stats_teams}
\end{table}

To capture team strengths which are relevant in predicting match outcomes, Table~\ref{tab:stats_teams} presents summary statistics on market values (sourced from \url{www.transfermarkt.com}) and Elo ratings as of 1 September 2024 and 1 September 2025, respectively, for both the UCL and UEL, separated by seeding pots and aggregated across all clubs. Elo ratings are obtained from \url{www.clubelo.com}. This variant of Elo ratings has also been used by, e.g.\ \citet{bosker2024impact}; for an overview of individual Elo ratings for UCL clubs, see Table~\ref{tab:elo} in Appendix~C. The average market value of UCL participants is more than twice that of UEL clubs in both seasons. In both competitions, market values are strongly right-skewed, indicating that a small number of clubs possess exceptionally high market values, suggesting competitive imbalance. This pattern is particularly evident for Pots 1 and 2. Moreover, the variation in market values, both within the same pot and across all clubs, is considerably greater in the UCL than in the UEL, suggesting a lower degree of competitive balance in the UCL. For the Elo ratings, summary statistics are generally higher in the UCL than in the UEL, except for the minimum in Pot 4 in the 2025/26 season and the standard deviation.

\subsection{Summary statistics on match outcomes in the 2024/25 and 2025/26 UCL and UEL}
We examine the 288 matches played per competition during the league phases of both the UCL and UEL in the 2024/25 and 2025/26 seasons. Table~\ref{tab:stats_UL} provides summary statistics on match outcomes, including the average number of goals and the share of wins with a margin of at least four goals (``lopsided matches''). The appearance of a greater competitive imbalance in the UCL is supported by the higher average number of goals in the UCL (3.26 and 3.38) compared to only 2.83 and 2.68 in the UEL \citep{scarf2022skill}. Further evidence comes from the fact that 15.97\% (13.19\%) of UCL matches ended with one team winning by four goals or more, whereas only 3.47\% (4.17\%) of UEL matches produced such lopsided outcomes. While in both competitions and seasons about 50\% of the matches resulted in a home win, the frequency of away wins in the UEL is particularly lower in the 2024/25 season (difference from the UCL of approximately seven percentage points). Consequently, the proportion of draws in the UEL is nearly twice as high in that season. Taken together, these findings indicate a greater imbalance in the UCL, consistent with summary statistics presented in Table~\ref{tab:stats_teams}.

\begin{table}[htp!]
    \centering
    \scalebox{0.75}{
    \begin{tabular}{c|ccc|ccc|c}
         & \multicolumn{3}{c|}{Outcome} & \multicolumn{3}{c|}{Average goals} & Lopsided \\
         & Home win & Draw & Away win & Home & Away & Total & matches \\
        \hline
        UCL 2024/25 & 53.47\% & 12.50\% & 34.03\% & 1.88 & 1.39 & 3.26 & 15.97\%\\
        UCL 2025/26 & 49.31\% & 17.36\% & 33.33\% & 1.93 & 1.45 & 3.38 & 13.19\%\\
        \hline
        UEL 2024/25 & 48.61\% & 24.31\% & 27.08\% & 1.64 & 1.19 & 2.83 & 3.47\%\\
        UEL 2025/26 & 52.78\% & 17.36\% & 29.86\% & 1.52 & 1.16 & 2.68 & 4.17\%\\
    \end{tabular}}
    \caption{Summary statistics on the distribution of wins, average number of goals, and the percentage of ``lopsided matches'' with a difference of at least four goals for the UCL and UEL in the 2024/25 and 2025/26 seasons.}
    \label{tab:stats_UL}
\end{table}

\subsection{Summary statistics on previous UEFA competitions and national leagues}

To place the UCL and UEL seasons under the new format into context, we calculate the same statistics for the 2020/21 -- 2023/24 UCL seasons (see Table~\ref{tab:stats_CL_prev}; note that the distribution of results in the 2020/21 season is potentially affected by the COVID-19 pandemic and matches played behind closed doors, cf.\ \citealp{winkelmann2021bookmakers}) as well as for the 2023/24 seasons of the top four European domestic leagues (see Table~\ref{tab:stats_NL}), namely, the English Premier League, the German Bundesliga, the Italian Serie A, and the Spanish La Liga. While, on average, the national leagues (26.42\%) and the 2024/25 UEL season (24.31\%) recorded roughly twice as many draws as the 2024/25 UCL season (12.50\%), the proportion of draws in the last UCL seasons under the old format (20.05\%) was already lower but still substantially higher than in the 2024/25 UCL season. The pattern also holds for the 2025/26 season, where both competitions exhibit the same proportion of draws (17.36\%), substantially lower than in all national leagues. This points to a lower degree of competitive balance, particularly in the UCL, compared with national leagues. A similar pattern emerges when considering the proportion of lopsided matches, i.e.\ wins by a margin of at least four goals. Such outcomes occur only rarely in the UEL -- less frequently than in any of the national leagues -- whereas their share under the new UCL format (about 15\%) is more than double the average across national leagues (6.57\%). However, we also observe many lopsided matches for most previous UCL seasons.

\begin{table}[htp!]
    \centering
    \scalebox{0.75}{
    \begin{tabular}{c|ccc|ccc|c}
         & \multicolumn{3}{c|}{Outcome} & \multicolumn{3}{c|}{Average goals} & Lopsided\\
        League & Home win & Draw & Away win & Home & Away & Total & matches \\
        \hline
        UCL 2020/21 & 41.67\% & 20.83\% & 37.50\% & 1.52 & 1.49 & 3.01 & 12.50\% \\
        UCL 2021/22 & 48.96\% & 18.75\% & 32.29\% & 1.71 & 1.39 & 3.09 & 13.54\% \\
        UCL 2022/23 & 47.92\% & 19.79\% & 32.29\% & 1.68 & 1.49 & 3.17 & 16.67\% \\
        UCL 2023/24 & 46.88\% & 20.83\% & 32.29\% & 1.76 & 1.32 & 3.08 & 4.17\% \\
        \hline
        \textbf{UCL 2020/21 -- 2023/24} & \textbf{46.35\%} & \textbf{20.05\%} & \textbf{33.59\%} & \textbf{1.67} & \textbf{1.42} & \textbf{3.09} & \textbf{11.72\%} \\
        \hline
        UCL 2024/25 & 53.47\% & 12.50\% & 34.03\% & 1.88 & 1.39 & 3.26 & 15.97\%\\
        UCL 2025/26 & 49.31\% & 17.36\% & 33.33\% & 1.93 & 1.45 & 3.38 & 13.19\%\\

    \end{tabular}}
    \caption{Summary statistics on the distribution of wins, average number of goals, and the percentage of ``lopsided matches'' with a difference of at least four goals for the UCL seasons 2020/21 -- 2025/26.}
    \label{tab:stats_CL_prev}
\end{table}

Another important finding concerns home advantage. In the 2024/25 UCL season, the share of home wins is about ten percentage points higher than the average across national leagues, and still seven percentage points above the average of previous UCL seasons, consistent with earlier research on home advantage in the UCL \citep{kuvvetli2024home}. These differences are smaller but still positive for the 2025/26 season. The total number of goals in the UCL is approximately 14\% higher, driven primarily by an increase of nearly 18\% in goals scored by home teams, while away teams score only 9\% more goals than in national leagues. Relative to earlier UCL seasons, both home and total goals are higher in the 2024/25 and 2025/26 seasons.

\begin{table}[htp!]
    \centering
    \scalebox{0.75}{
    \begin{tabular}{c|ccc|ccc|c}
         & \multicolumn{3}{c|}{Outcome} & \multicolumn{3}{c|}{Average goals} & Lopsided\\
        League & Home win & Draw & Away win & Home & Away & Total & matches \\
        \hline
        Premier League 23/24 & 46.05\% & 21.58\% & 32.37\% & 1.80 & 1.48 & 3.28 & 8.68\%\\
        Bundesliga 23/24 & 43.79\% & 26.47\% & 29.74\% & 1.81 & 1.41 & 3.22 & 8.82\%\\
        Serie A 23/24 & 41.84\% & 29.47\% & 28.68\% & 1.43 & 1.18 & 2.61 & 4.47\%\\
        La Liga 23/24 & 43.95\% & 28.16\% & 27.89\% & 1.48 & 1.16 & 2.64 & 4.47\%\\
        \hline
        All & 43.91\% & 26.42\% & 29.67\% & 1.62 & 1.30 & 2.92 & 6.57\%\\
    \end{tabular}}
    \caption{Summary statistics on the distribution of wins, average number of goals, and the percentage of ``lopsided matches'' with a difference of at least four goals for the top four European national leagues in the 2023/24 season.}
    \label{tab:stats_NL}
\end{table}

From this descriptive analysis, we can conclude that the distribution of team strengths in the UCL is more unequal than in both the national leagues and the UEL. This imbalance is reflected in a higher proportion of (lopsided) wins, particularly for home teams (approximately 70\% of lopsided wins are achieved by home teams), and in fewer draws. Specifically, the 2024/25 UCL exhibits a markedly lower share of draws than observed in the UEL, previous UCL seasons, or the national leagues. This result may be driven by stronger incentives to play for a win rather than to settle for a draw, as already outlined in the Introduction. The frequency of draws in the UCL increases in the 2025/26 season, but remains clearly below that of national leagues and UCL seasons under the old format. At the same time, we observe fewer draws in the second UEL season under the new format, equal to the UCL frequency.

\section{Modelling approach}
\label{sec:prediction}
To model match outcomes and subsequently predict qualification thresholds, we employ two bivariate Poisson models. First, in Section~\ref{sec:indpois}, we introduce the independent version (\citealp{maher1982modelling}). Second, in Section~\ref{sec:dc}, we present one of its extensions, the bivariate Poisson model proposed by \citet{dixon1997modelling}. Afterwards, in Section~\ref{sec:predictor}, we outline the explicit form of the linear predictors.

\subsection{Independent bivariate Poisson model}\label{sec:indpois}

Let $X$ and $Y$ be random variables for the goals scored by the home and away team. Under the basic model, the probability of observing a particular scoreline $(x,y)$ is given by

\begin{equation}
P(X=x, Y=y) = \frac{\lambda^x e^{-\lambda}}{x!} \cdot \frac{\mu^y e^{-\mu}}{y!},
\end{equation}
where $\lambda$ and $\mu$ are the expected number of goals for the home and away team, respectively. This formulation implicitly assumes independence between the home and away goal distributions.

\subsection{Dixon--Coles model}\label{sec:dc}
Again, let $X$ and $Y$ denote the goals scored by both teams. Under the model proposed by \citet{dixon1997modelling}, the probability of observing a particular scoreline $(x,y)$ is given by

\begin{equation}
P(X=x, Y=y) = \tau_{\rho}(x,y) \cdot \frac{\lambda^x e^{-\lambda}}{x!} \cdot \frac{\mu^y e^{-\mu}}{y!},
\end{equation}
where $\lambda$ and $\mu$ are the expected number of goals for the home and away team, and $\tau_{\rho}(x,y)$ is a correction factor applied to low-scoring outcomes (0-0, 0-1, 1-0, 1-1), defined as

\begin{equation}
\tau_{\rho}(x,y) = 
\begin{cases} 
1 - \lambda \mu \rho & \text{if } x=y=0,\\
1 + \lambda \rho & \text{if } x=0, y=1,\\
1 + \mu \rho & \text{if } x=1, y=0,\\
1 - \rho & \text{if } x=y=1,\\
1 & \text{otherwise},
\end{cases}
\end{equation}
where $\rho$ is a dependence parameter that decides whether probabilities shift in one or the other direction. While the Dixon--Coles model was initially developed to shift probability mass from 1–0 and 0–1 outcomes towards 0–0 and 1–1 outcomes ($\rho < 0$), \citet{michels2025extending} show that it can also redistribute probabilities in the opposite direction ($\rho > 0$). This feature is particularly relevant for our application, as the recent UCL seasons exhibited \textit{fewer} draws than suggested by an independent model. Another noteworthy property is that, for $\rho = 0$, the model reduces to the case of complete independence between home and away scores, i.e.\ the classical independent bivariate Poisson model introduced in Section~\ref{sec:indpois}.

\subsection{Predictor modelling}\label{sec:predictor}
While Opta relied on their \textit{Power Rankings} and betting odds, we follow the approach suggested in the literature and proxy team-specific strengths using Elo ratings \citep{hvattum2010using}. This choice is motivated by the need to forecast matches between previously unseen teams, which renders team-fixed effects (as employed, for example, by \citealp{dixon1997modelling} and \citealp{otting2024demand}) infeasible. While betting odds could also serve as proxies \citep{michels2023bettors}, they are typically only available shortly before matches and do not exist for randomly generated match schedules considered in our sensitivity analyses. Elo ratings, by contrast, represent a dynamic measure of team strength based on previous match results and the observed outcome given the team's strength. They can be interpreted as a relative indicator of competitive performance on a common scale across leagues, making them particularly suitable for European tournament settings \citep{csato2024club}.

Thus, we model the expected number of home goals $\lambda$ and away goals $\mu$ (scoring intensities) as a log‑linear function of the Elo rating difference, denoted as \textit{EloDiff}. Thereby, we treat Elo as a continuous proxy for relative team strength within a standard Poisson framework. While some studies rely on the Elo win‑probability formula (see, e.g.\ \citealp{csato2025competitive}), our specification avoids imposing this non‑linear transformation and allows the relationship between rating differences and goal scoring to be determined empirically.
\begin{align*}
\lambda &= \exp\left(\beta_0 + \beta_1 \cdot \text{EloDiff} + \beta_2 \right),\\
\mu &= \exp\left(\beta_0 - \beta_1 \cdot \text{EloDiff} \right),
\label{eq:mod}
\end{align*}
where $\beta_0, \beta_1, \beta_2$ are parameters to be estimated, including the home effect $\beta_2$. The sign of the term that includes $\text{\textit{EloDiff}} = \text{\textit{Elo}}_\text{\textit{home}} - \text{\textit{Elo}}_\text{\textit{away}}$ differs due to the perspective of the home and away team, respectively. This parametrisation enables the model to directly incorporate differences in team strength as measured by Elo ratings.

\section{Results}
\label{sec:equal}
In this section, we provide simulations of the final standings and the points required to proceed to the knockout stage. First, we outline the simulation setup (Section~\ref{sec:setup}). We then fit the independent bivariate Poisson model from Section~\ref{sec:indpois} to data from the 2023/24 season of top European domestic leagues in order to mimic the Opta forecast for the 2024/25 UCL season under the official match schedule (Section~\ref{sec:pred_national}). Afterwards, in Section~\ref{sec:predict_2526}, we apply the Dixon--Coles model from Section~\ref{sec:dc} to data from the 2024/25 UCL and UEL seasons and predict the 2025/26 seasons. Subsequently, we provide model checks and extend our out-of-sample analysis by randomly generating match schedules to examine whether the results are driven by the official schedule, enabling more reliable predictions of the point thresholds in both competitions (Section~\ref{sec:sim}). Finally, in Section~\ref{sec:sensitivity}, we perform sensitivity analyses.

\subsection{Simulation setup}
\label{sec:setup}
To estimate the number of points required to progress to the play-offs and the Round of 16, in each simulation experiment, we use a match schedule of 36 teams allocated to four seeding pots. Each club plays against two opponents from each pot, with one match at home and one away. Seeding pots and schedules are either given by the official schedules for the 2024/25 season (Section~\ref{sec:pred_national}), the 2025/26 season (Section~\ref{sec:predict_2526}), or randomly generated schedules (Sections~\ref{sec:sim} and~\ref{sec:sensitivity}). Outcome probabilities for each match are derived from the underlying regression model.

In each simulation run, we randomly draw an outcome for each match according to the underlying probability distribution. A win awards three points, a draw one point to each team, and a loss zero points. After simulating all matches, the teams are ranked in descending order according to the points accumulated. Cut-off values for qualification, i.e.\ the lowest number of points sufficient to advance, to the Round of 16 and the play-offs are then determined. For each team in each simulation run, qualification is coded as 1, and non-qualification as 0. In cases where multiple teams finish with the same number of points but not all progress to the next round, we calculate qualification chances as $1/n$, where $n$ is the number of teams with the same points total at the qualification threshold. We repeat each simulation 10,000 times. Averaging across all runs yields an estimate of the likelihood of qualification for each possible points total. Note that, for simplicity, we do not account for the exact match schedule, as our model relies on fixed Elo ratings from the start of the season.

\subsection{Predicting UEFA competitions based on national league data}
\label{sec:pred_national}
As a first step, we use data from the top four European domestic leagues to fit an independent bivariate Poisson model based on Elo ratings, as introduced in Sections~\ref{sec:indpois} and~\ref{sec:predictor}. For simplicity, we use static Elo ratings as of 1 September 2023. The estimated coefficients are reported in Table~\ref{tab:mod1} together with corresponding standard errors. As expected, a higher (lower) difference in Elo ratings between the observed team and its opponent increases (decreases) the average goals of the observed team. Furthermore, playing at home increases the team's average number of goals. For a balanced match with average Elo ratings of 1,405 for both teams, the model predicts, on average, $1.56$ goals for the home team and $1.25$ goals for the away team. 

\begin{table}[!htbp] \centering 
  \caption{Estimated coefficients for the independent bivariate Poisson model based on national league data.} 
  \label{tab:mod1} 
  \scalebox{0.75}{
\begin{tabular}{@{\extracolsep{5pt}}lc} 
\\[-1.8ex]\hline 
\hline \\[-1.8ex] 
 & \multicolumn{1}{c}{\textit{Dependent variable:}} \\ 
\cline{2-2} 
\\[-1.8ex] & Goals \\ 
\hline \\[-1.8ex] 
 EloDiff & 0.0019$^{***}$ \\ 
  & (0.0001) \\ 
  & \\ 
 Home & 0.220$^{***}$ \\ 
  & (0.031) \\ 
  & \\ 
 Constant & 0.225$^{***}$ \\ 
  & (0.023) \\ 
  & \\ 
\hline 
Observations & 1446 \\
\hline
\hline \\[-1.8ex] 
\textit{Note:}  & \multicolumn{1}{r}{$^{*}$p$<$0.1; $^{**}$p$<$0.05; $^{***}$p$<$0.01} \\ 
\end{tabular}}
\end{table}
We use these estimates to simulate the number of goals from two independent Poisson distributions and determine the match outcome in each simulation run. By conducting a sufficiently large number of runs, we can estimate the win probability distribution for each match based on the relative frequencies of simulated outcomes. For the example given above of teams with identical Elo ratings, we obtain probabilities of 44.3\% for a home win, 24.9\% for a draw, and 30.8\% for an away win.

\begin{figure}[htp!]
    \centering
    \subfigure[UCL]{\includegraphics[width=0.5\textwidth]{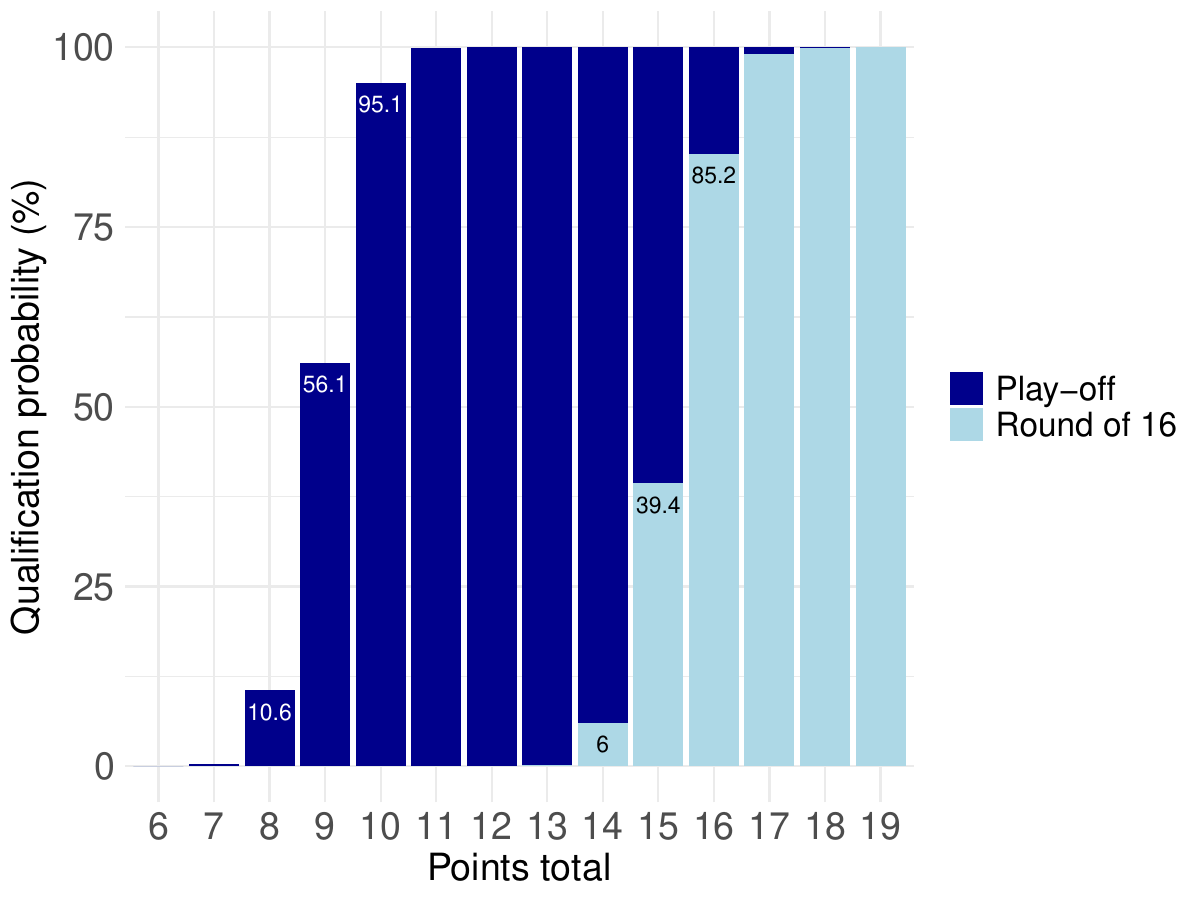}}
    \subfigure[UEL]
    {\includegraphics[width=0.48\textwidth]{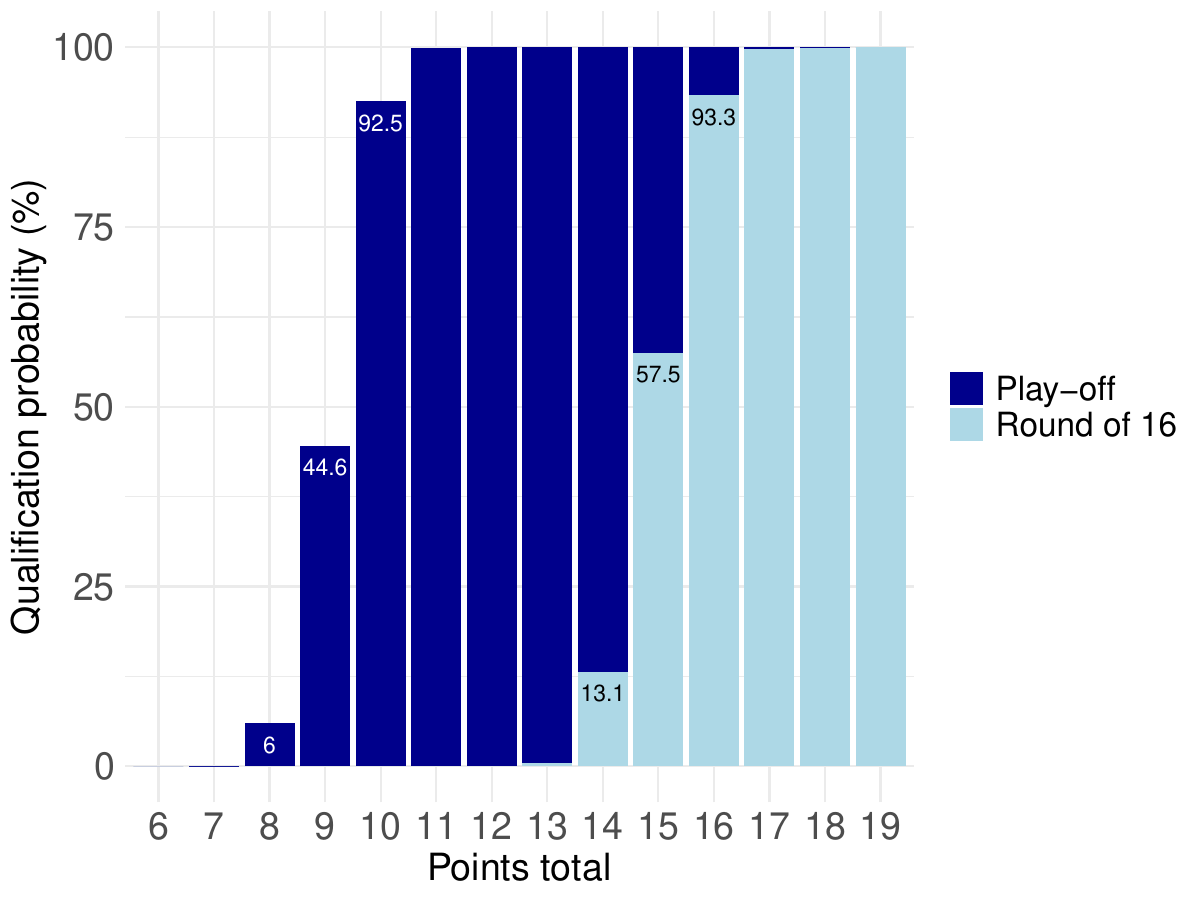}}
    \caption{Probability to progress to the Round of 16 and play-offs, respectively, for the 2024/25 UCL and UEL season schedules, depending on the number of points achieved using a prediction model trained on national league data (Table~\ref{tab:mod1}).}
    \label{fig:clel_2425}
\end{figure}

We now use the estimated coefficients from Table~\ref{tab:mod1} to predict match outcomes for the 2024/25 UCL and UEL season based on the official match schedule and Elo ratings for clubs. The procedure follows the approach outlined in Section~\ref{sec:setup}. Figure~\ref{fig:clel_2425} presents the resulting qualification thresholds for the Round of 16 and play-offs in both competitions. Comparing our findings to the pre-competition forecasts by Opta based on their Power Rankings and betting odds (see Figure~\ref{fig:opta} in the Introduction), we observe some differences. For the UCL, both approaches indicate a probability close to 100\% of qualifying for the Round of 16 with 17 points or more. However, for lower point totals, the predictions diverge: Opta suggests probabilities of 73\% (28\%) for 15 (14) points, whereas our approach, using domestic league data, yields only 39.4\% (6.0\%). Similarly, for qualification to the play-off round, Opta suggests probabilities of 69\% (16\%) for 9 (8) points, compared with our values of 56.1\% (10.6\%).\\

Likely driven by the difference in competitive balance between the UCL and UEL (see Section~\ref{sec:data}), we observe corresponding variations in the qualification thresholds. For the Round of 16, qualification is slightly more likely in the UEL than in the UCL for the same number of points, indicating that a higher point total is required to advance in the UCL. Specifically, the qualification probability in the UEL exceeds that in the UCL by 18 percentage points at 15 points and by seven percentage points at 14 points. Notably, we find a reverse result for the play-off round: Qualification probabilities in the UEL are lower by up to 12 percentage points (at nine points) compared with the UCL. The requirement of higher point totals to reach the Round of 16 in the UCL can potentially be explained by the greater competitive imbalance at the top (see Table~\ref{tab:stats_teams} in Section~\ref{sec:descriptive}) and the resulting lower number of draws compared with the UEL. In contrast, teams in the lower part of the UCL table collect fewer points than in the UEL, which leads to slightly higher qualification probabilities for the play-off round for a given point total in the UCL, compared to the UEL.

In conclusion, the predicted threshold point totals for both competitions are similar to those reported by Opta. Slightly lower qualification probabilities can be explained by our use of Elo ratings. Nevertheless, the predicted thresholds still do not align with the number of points required to advance to the upcoming rounds in the 2024/25 season, particularly in the UCL. We suggest that the main reason for this discrepancy is that the independent bivariate Poisson model predicts, on average, 31 draws for the UCL -- approximately 70\% higher than the 18 draws actually observed during the 2024/25 season. Indeed, only in 46 out of 10,000 simulation runs does the model suggest 18 or fewer draws.

\subsection{Predicting UEFA competitions with the Dixon--Coles Model}
\label{sec:predict_2526}

Given the results of the first season under the new competition, and our findings from Section~\ref{sec:pred_national}, we now aim to enhance the prediction accuracy of qualification thresholds in the UCL and UEL league phases. For this purpose, we replace the independent bivariate Poisson model with the model suggested by \citet{dixon1997modelling}. This approach specifically allows probabilities to be shifted from draws to home and away wins, thereby accounting for the relatively low number of draws observed in the 2024/25 UCL season (see Section~\ref{sec:dc}). We estimate the model based on this data and (1) conduct an in-sample analysis using the official 2024/25 match schedule, comparing results to those obtained in Section~\ref{sec:pred_national}, and (2) perform out-of-sample simulations using the 2025/26 match schedule. Moreover, in Appendix~\ref{sec:app_comparison}, we conduct model comparisons using different covariates and proper scoring rules \citep{gneiting2007strictly}.

\begin{table}[!htbp] \centering 
  \caption{Estimated coefficients for the Dixon--Coles models fitted to the 2024/25 UCL and UEL season.} 
  \label{tab:mod_dc_2425} 
  \scalebox{0.75}{
\begin{tabular}{@{\extracolsep{5pt}}lcc} 
\\[-1.8ex]\hline 
\hline \\[-1.8ex] 
 & \multicolumn{2}{c}{\textit{Dependent variable: Goals}} \\ 
\cline{2-3} 
\\[-1.8ex] & UCL & UEL \\ 
\hline \\[-1.8ex] 
 EloDiff & 0.0021$^{***}$ & 0.0013$^{***}$ \\ 
  & (0.0002) & (0.0003)\\ 
  & \\ 
 Home & 0.296$^{***}$ & 0.317$^{***}$ \\ 
  & (0.094) & (0.100)\\ 
  & \\ 
 $\rho$ & 0.102 & 0.005 \\ 
  & (0.096) & (0.111)\\ 
  & \\ 
 Constant & 0.246$^{***}$ & 0.157$^{**}$ \\ 
  & (0.073) & (0.077)\\ 
  & \\ 
\hline 
Observations & 144 & 144 \\
\hline
\hline \\[-1.8ex] 
\textit{Note:}  & \multicolumn{1}{r}{$^{*}$p$<$0.1; $^{**}$p$<$0.05; $^{***}$p$<$0.01} \\ 
\end{tabular} }
\end{table}

Table~\ref{tab:mod_dc_2425} provides parameter estimates for the models fitted to data from the 2024/25 UCL (left column) and UEL (right column) seasons, together with corresponding standard errors. We find a stronger effect of the difference in Elo ratings in the UCL, suggesting a higher likelihood of (lopsided) wins in more imbalanced matches than in the UEL. The estimated correlation parameter, which governs the redistribution of probability mass from draws to wins, is denoted by $\hat{\rho}$. It is estimated to be close to zero in the UEL, implying that the more complex model offers little advantage over the independent bivariate Poisson model. In contrast, for the UCL, we obtain $\hat{\rho} = 0.102$, indicating that, for teams with average Elo ratings, the probability of a 1:1 is reduced by 10.2\%, relative to the independent model, and that of a 0:0 draw by 22.4\%. Given that the model includes only one season under the incomplete round-robin format, yielding 144 observations per competition, it is unsurprising that the effect is statistically insignificant even for the UCL.

\begin{figure}[htp!]
    \centering
    \subfigure[UCL]{\includegraphics[width=0.48\textwidth]{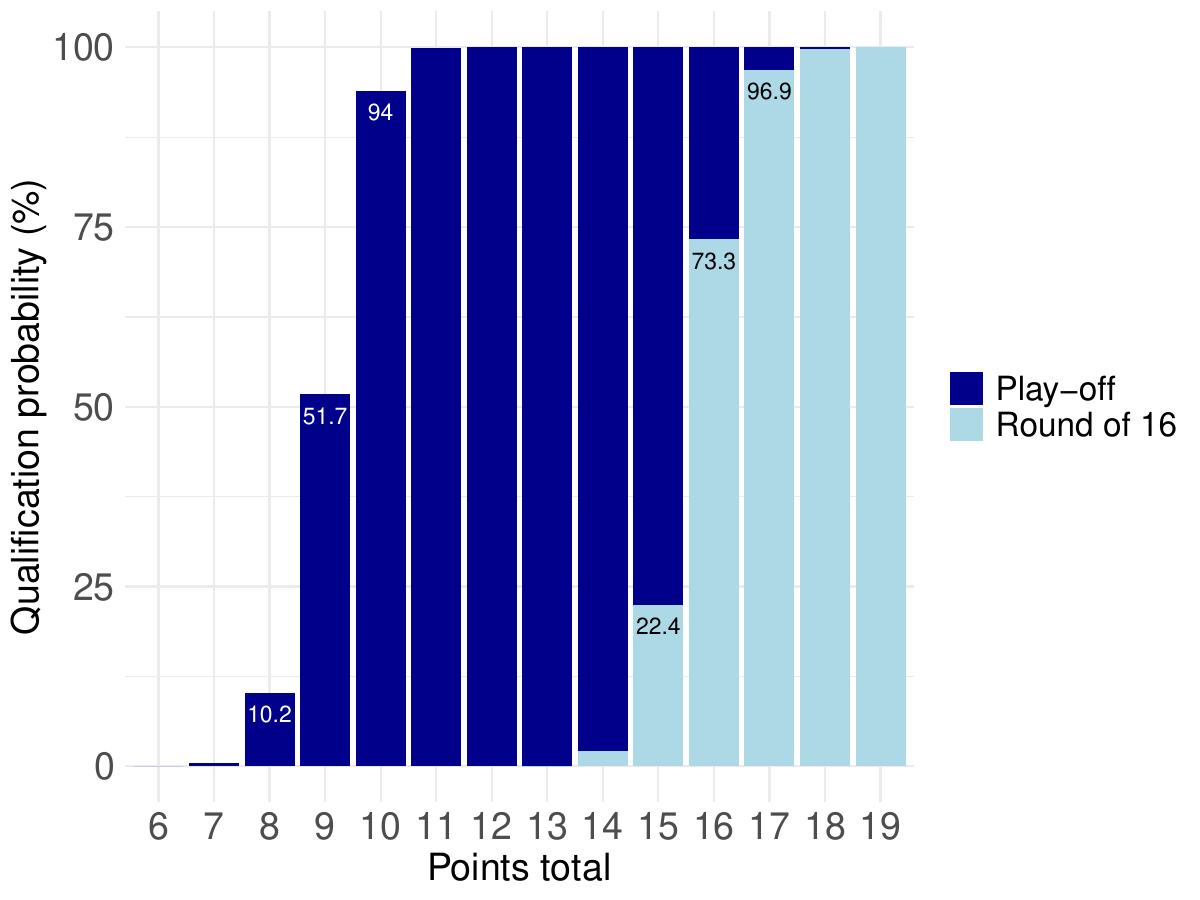}}
    \subfigure[UEL]
    {\includegraphics[width=0.48\textwidth]{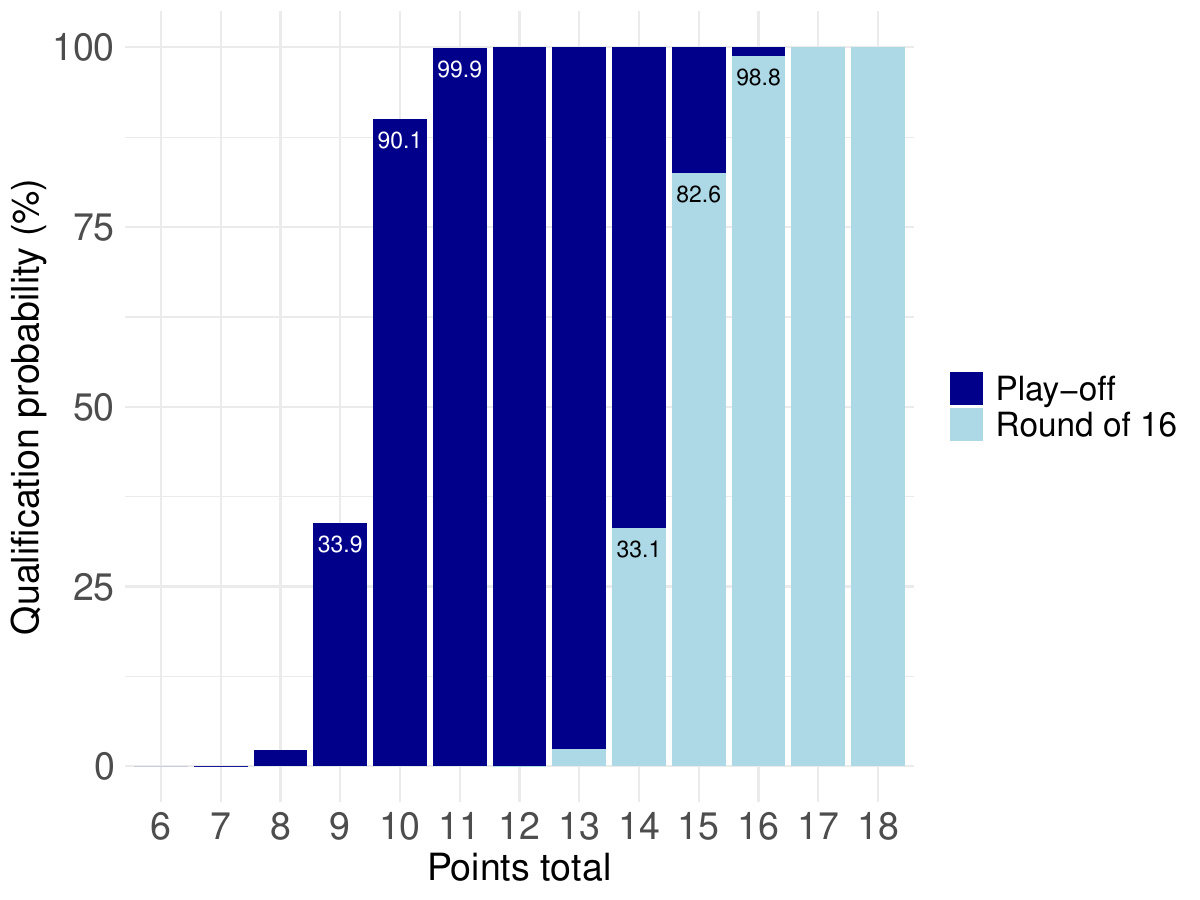}}
    \caption{Probability to progress to the Round of 16 and play-offs, respectively, in the 2024/25 season (in-sample simulation), depending on the number of points achieved. The prediction is based on the win probabilities estimated from the 2024/25 season schedule with the Dixon--Coles model (Table~\ref{tab:mod_dc_2425}).}
    \label{fig:clel_dc_2425}
\end{figure}

We first use the estimated coefficients for an in-sample simulation to predict qualification thresholds for the 2024/25 season. The results are presented in Figure~\ref{fig:clel_dc_2425}. For the UCL, the model produces, on average, only 26 draws -- nearly 20\% fewer than the model trained on domestic league data, and closer to the 18 draws observed empirically (18 or fewer draws are produced in 563 simulation runs). Consequently, qualification thresholds for specific point totals change substantially. For instance, the probability of directly reaching the Round of 16 with 16 points decreases from 85.2\% to 73.3\%. Similar effects are observed for 15 (14) points, where probabilities decline from 39.4\% (6.0\%) to 22.4\% (2.2\%). For the UEL, however, the effect is reversed: qualification probabilities increase markedly, from 57.5\% (13.1\%) to 82.6\% (33.1\%) for 15 (14) points. Indeed, empirical results for the first season show that all teams with 15 points advanced directly to the Round of 16 in the UEL, whereas all teams collecting 15 points in the UCL had to compete in the play-off round. Regarding progression to the play-offs, the two approaches yield only minor differences in qualification probabilities for the UCL. For the UEL, however, the likelihood of reaching the play-off round with nine points decreases by about 11 percentage points from 44.6\% to 33.9\%. In the 2024/25 UEL season, both teams finishing with nine points were eliminated.

\begin{figure}[htp!]
    \centering
    \subfigure[UCL]{\includegraphics[width=0.48\textwidth]{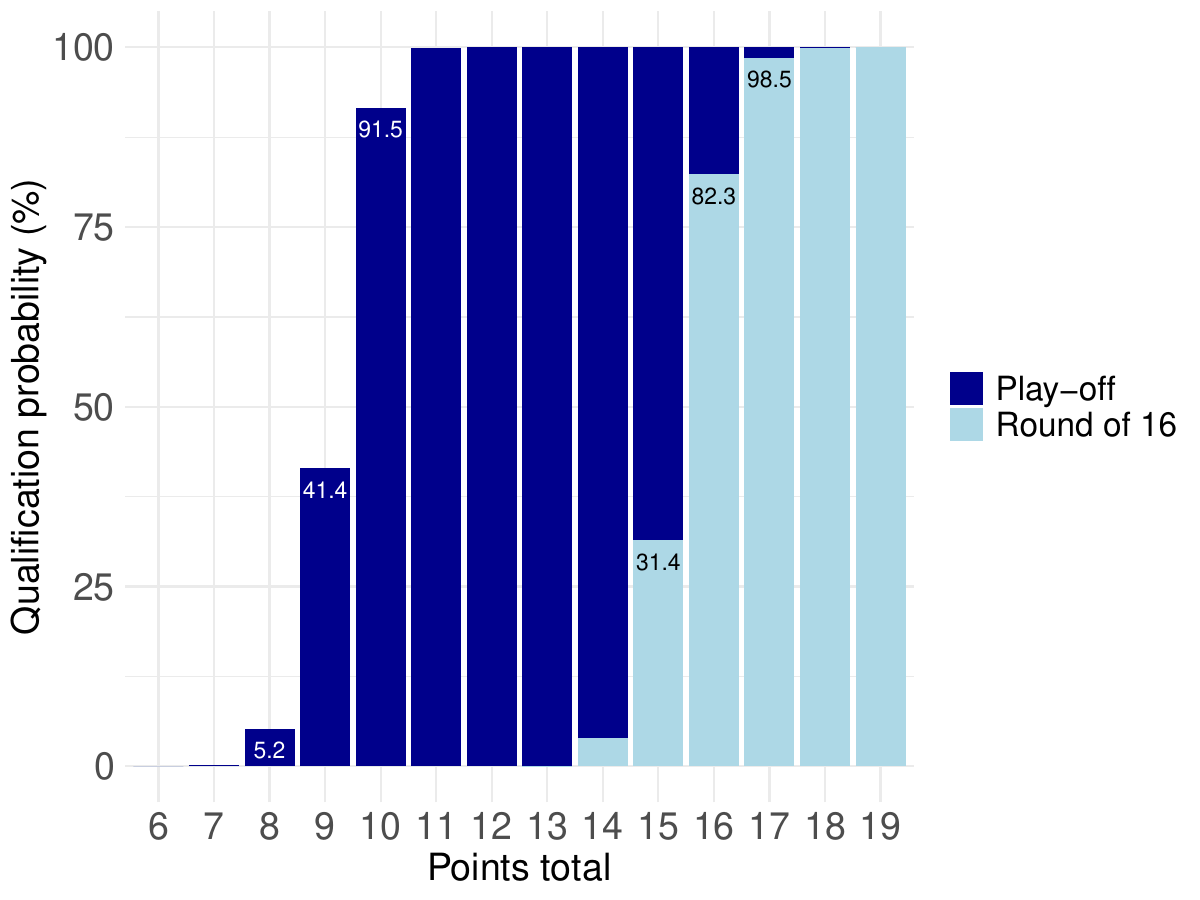}}
    \subfigure[UEL]
    {\includegraphics[width=0.48\textwidth]{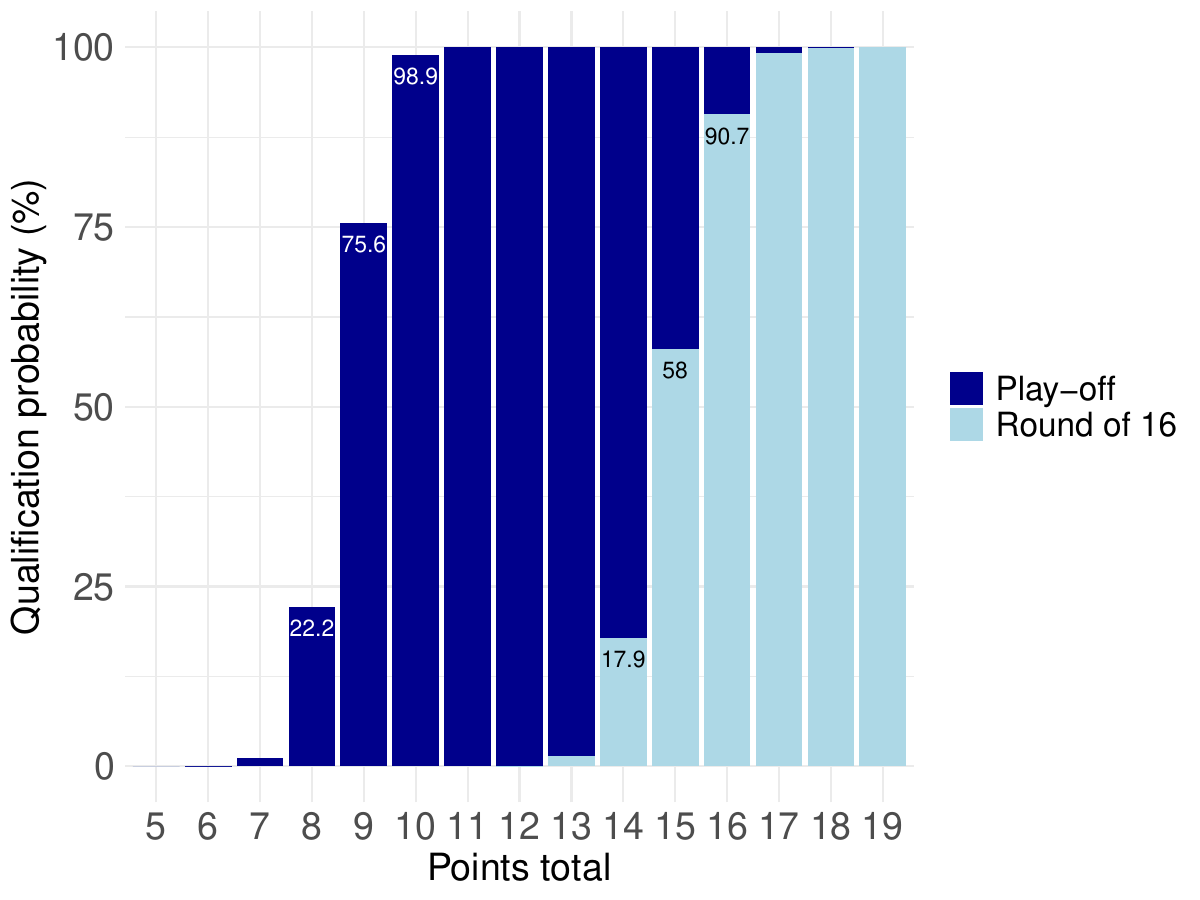}}
    \caption{Probability to progress to the Round of 16 and play-offs, respectively, in the 2025/26 season (out-of-sample simulation), depending on the number of points achieved. The prediction is based on the win probabilities estimated from the 2024/25 season schedule with the Dixon--Coles model (Table~\ref{tab:mod_dc_2425}).}
    \label{fig:clel_dc_2526}
\end{figure}

In the following, we employ the same estimation model and apply it to the official match schedule from the 2025/26 season for two main reasons. First, we aim to evaluate the predictive power of our approach using an out-of-sample dataset. Second, we seek to investigate whether the composition of participating clubs and the match schedule affect qualification thresholds. Our model predicts, on average, 26 draws in the UCL, which is fairly close to the 25 draws observed empirically (4,904 simulation runs produce at most 25 draws). Moreover, Figure~\ref{fig:clel_dc_2526} illustrates the predicted qualification thresholds. When comparing these thresholds with those from the previous season, we observe differences of up to ten percentage points in the UCL. While the probability of qualifying for the play-offs with a given points total is higher in the 2024/25 season, the probability of reaching the Round of 16 is higher in the 2025/26 season. For instance, the predicted qualification probability for 16 points is 73.3\% in the 2024/25 season, compared with 82.3\% in the 2025/26 season. In both seasons, all teams with 16 points qualified directly for the Round of 16.

We find a reverse pattern for the UEL, where teams with nine points had a 75.6\% probability of qualifying for the play-offs in the 2025/26 season (only 33.9\% in the 2024/25 season). However, the 58.0\% probability of directly advancing to the Round of 16 with 15 points in the 2025/26 season is clearly lower than the 82.6\% in the preceding season. This pattern is consistent with the empirical results. In the 2024/25 season, two of the five teams with ten points were eliminated, whereas in 2025/26 one team with only nine points reached the play-offs. Conversely, in 2024/25, one team with 14 points qualified directly for the Round of 16, while in 2025/26 even one team with 16 points had to take part in the play-offs. These findings confirm that the Dixon--Coles model can generate realistic qualification thresholds in an out-of-sample dataset (for model checks and comparisons, see Appendix~\ref{sec:app_comparison}). Nevertheless, the differences between the two seasons demonstrate that the composition of participating teams affects qualification thresholds.

\subsection{Simulating UEFA competition results}
\label{sec:sim}

As our results in the previous section indicate that the composition of clubs participating in the competition affects qualification thresholds, we now aim to generalise our findings. To this end, we first estimate the Dixon--Coles model using data pooled from both seasons. Table~\ref{tab:mod_dc_combined} presents results for the Dixon--Coles model fitted to the combined dataset. Compared with the model estimated for the 2024/25 season alone, we observe slight differences across all estimated coefficients. In particular, $\hat{\rho}$ decreases from 0.102 to 0.036. This change is plausible, given the higher number of draws in the 2025/26 season compared with the previous one. Nevertheless, the positive value still indicates a probability shift from draws towards wins. As expected, the larger sample size results in smaller standard errors, yet the conclusions regarding statistical significance remain unchanged.

\begin{table}[!htbp] \centering 
  \caption{Estimated coefficients for the Dixon--Coles models fitted to the 2024/25 and 2025/26 UCL and UEL seasons.} 
  \label{tab:mod_dc_combined} 
  \scalebox{0.75}{
\begin{tabular}{@{\extracolsep{5pt}}lcc} 
\\[-1.8ex]\hline 
\hline \\[-1.8ex] 
 & \multicolumn{2}{c}{\textit{Dependent variable: Goals}} \\ 
\cline{2-3} 
\\[-1.8ex] & UCL & UEL \\ 
\hline \\[-1.8ex] 
 EloDiff & 0.0018$^{***}$ & 0.0015$^{***}$ \\ 
  & (0.0002) & (0.0002)\\ 
  & \\ 
 Home & 0.289$^{***}$ & 0.294$^{***}$ \\ 
  & (0.065) & (0.072)\\ 
  & \\ 
 $\rho$ & 0.036 & 0.064 \\ 
  & (0.071) & (0.078)\\ 
  & \\ 
 Constant & 0.286$^{***}$ & 0.133$^{**}$ \\ 
  & (0.051) & (0.055)\\ 
  & \\ 
\hline 
Observations & 288 & 288 \\
\hline
\hline \\[-1.8ex] 
\textit{Note:}  & \multicolumn{1}{r}{$^{*}$p$<$0.1; $^{**}$p$<$0.05; $^{***}$p$<$0.01} \\ 
\end{tabular} }
\end{table}

We now use these results to simulate 10,000 repetitions of the UCL and UEL. In each run, we randomly draw participating clubs and construct match schedules in accordance with the official draw procedure. To generate participating clubs, we combine those that took part in the 2024/25 and 2025/26 seasons, grouped by seeding pots. For each pot, we assume the Elo ratings of the clubs are uniformly and independently distributed between the minimum and maximum Elo ratings observed within that pot across both seasons. To perform draws under the new UCL competition format, \citet{devriesere2026groups} propose an integer programming model. As our analyses rely on fixed Elo ratings, the specific match sequence does not influence the results; we are only interested in the set of opponents assigned to each team. However, drawing procedures for group and league competitions are not trivial \citep{csato2025fairness}. Our simulation design ensures that every club faces exactly two teams from each seeding pot, one at home and one away. For simplicity, we do not account for association-related constraints, such as the restriction preventing clubs from the same association from meeting in the league phase. If a deadlock occurs in a single draw, the sample is discarded and redrawn. Finally, we predict match outcomes and qualification probabilities using the underlying prediction model.

\begin{figure}[htp!]
    \centering
    \subfigure[UCL]{\includegraphics[width=0.48\textwidth]{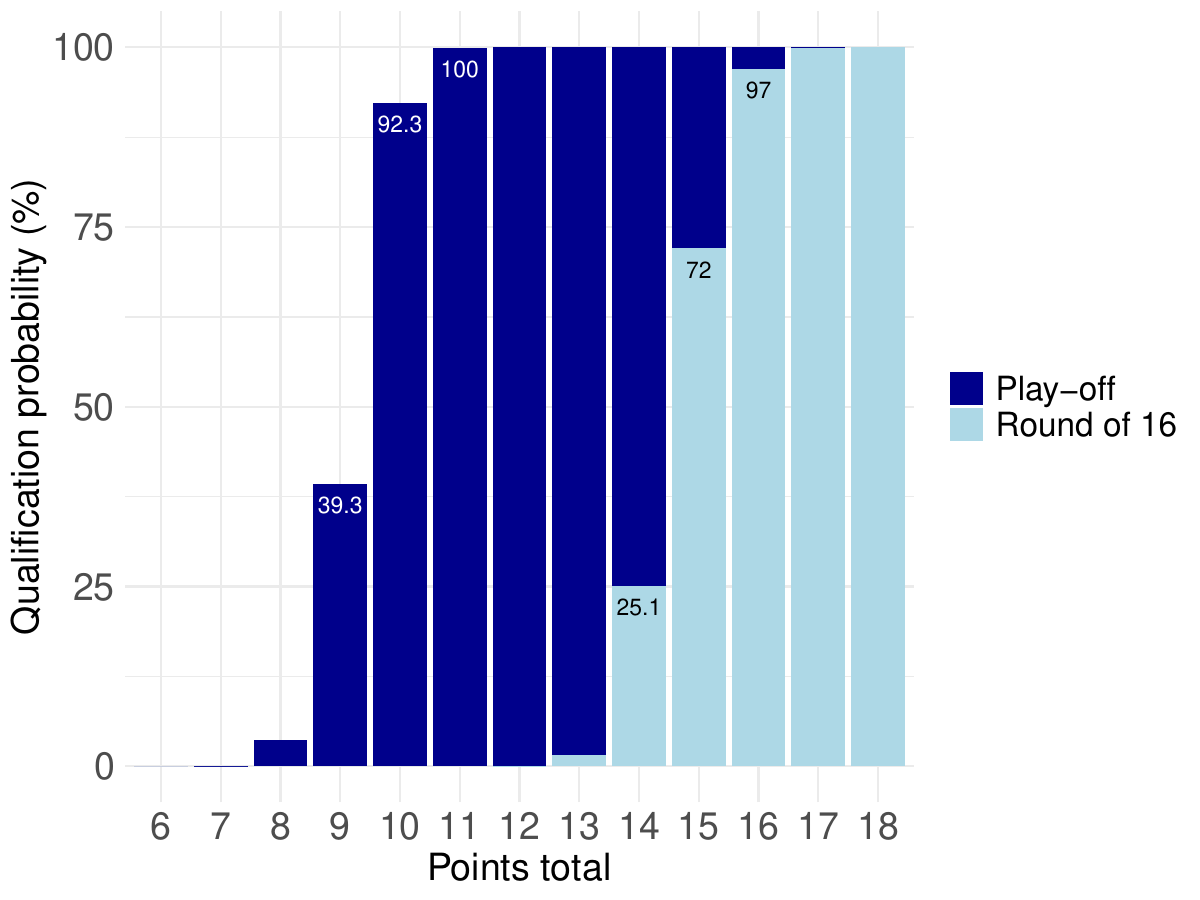}}
    \subfigure[UEL]
    {\includegraphics[width=0.48\textwidth]{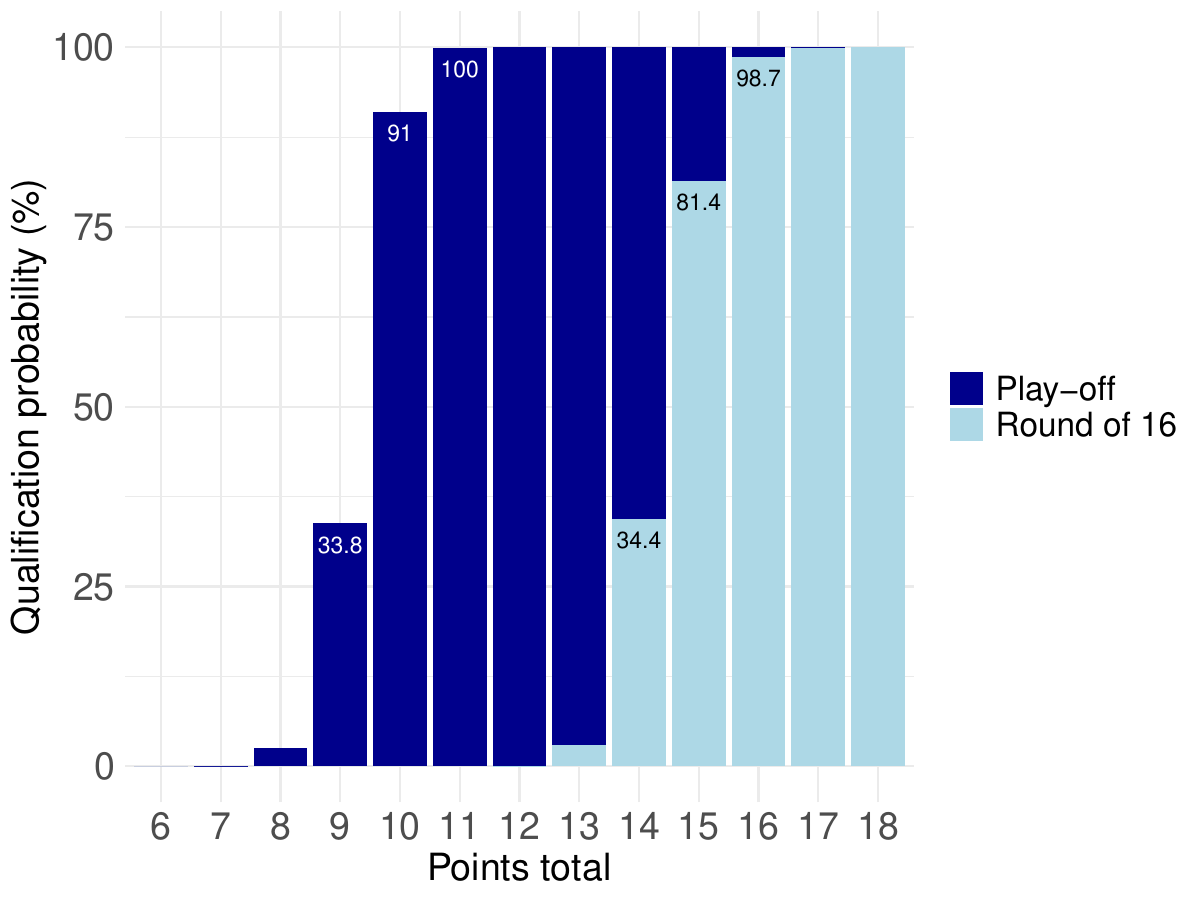}}
    \caption{Probability to progress to the Round of 16 and play-offs, respectively, in simulations with randomly sampled participating clubs and match schedules, depending on the number of points achieved. The prediction is based on the win probabilities estimated from the 2024/25 and 2025/26 seasons' schedule with the Dixon--Coles model (Table~\ref{tab:mod_dc_combined}).}
    \label{fig:clel_dc_2426}
\end{figure}

Figure~\ref{fig:clel_dc_2426} provides qualification thresholds for randomly drawn sets of participating clubs and match schedules. The results for the UCL and UEL are broadly similar, with only minor differences. For a given point total, qualification prospects are slightly higher in the UCL, by up to nine percentage points. Nevertheless, there is considerable variation across simulation runs. In individual simulation runs, seven points were sufficient to reach the play-off phase in both competitions, while in others, even 11 points led to elimination. The variation is even more striking for the Round of 16: in certain runs, 12 points sufficed, whereas in others, 18 (17) points still resulted in elimination in the UCL (UEL). Comparing these findings with the empirical observations from the first two seasons under the new system, the eliminations of VfB Stuttgart and GNK Dinamo Zagreb in the 2024/25 season, despite having collected ten and even 11 points, respectively, appear to have been rare events, largely driven by the exceptionally low number of draws in that season. According to our simulations, their qualification probabilities were 92.3\% (99.96\%).

\subsection{Sensitivity analysis}
\label{sec:sensitivity}
Given that the estimated coefficient for the parameter shifting probabilities from draws to wins, $\hat{\rho}$, is statistically insignificant, likely due to the limited number of observations, we conduct a sensitivity analysis to assess the effect of different values of $\rho$ on the qualification probabilities for specific point totals. In this analysis, we refer to the same procedure as in Section~\ref{sec:sim}, but vary $\rho \in [0, 0.2]$ while holding all other parameter values constant (see Table~\ref{tab:mod_dc_combined}).

\begin{table}[!htbp] \centering 
  \caption{Average number of draws observed in the simulation runs for the UCL (first column) and UEL (second column) depending on the parameter value $\rho$ for randomly generated sets of participating clubs and match schedules.}
  \label{tab:draws_sensitivity} 
  \scalebox{0.65}{
\begin{tabular}{@{\extracolsep{5pt}} cccccccccccc} 
\\[-1.8ex]\hline 
\hline \\[-1.8ex] 
Parameter value & $\rho = 0$ & $\rho = 0.02$ & $\rho = 0.04$ & $\rho = 0.06$ & $\rho = 0.08$ & $\rho = 0.10$ & $\rho = 0.12$ & $\rho = 0.14$ & $\rho = 0.16$ & $\rho = 0.18$ & $\rho = 0.2$ \\
\hline
UCL & 30.09 & 29.54 & 29.01 & 28.47 & 27.92 & 27.39 & 26.83 & 26.31 & 25.75 & 25.23 & 24.70 \\ 
UEL & 34.40 & 33.74 & 33.06 & 32.42 & 31.76 & 31.09 & 30.43 & 29.78 & 29.14 & 28.46 & 27.79\\
\hline \\[-1.8ex] 
\end{tabular}}
\end{table}

Table~\ref{tab:draws_sensitivity} reports the average number of draws observed in the simulation runs for the UCL and UEL, depending on the parameter value $\rho$. Even under an independent bivariate Poisson model ($\rho = 0$), we find more draws in the UEL compared with the UCL. As implied by the model properties, an increase in the value of $\rho$ shifts probability mass from draws to wins for either team. Accordingly, the average number of draws declines from 30.09 to 24.70 in the UCL and from 34.40 to 27.79 in the UEL when $\rho$ is increased to $0.2$. Recall that we estimated $\hat{\rho} = 0.036$ ($\hat{\rho} = 0.102$ for the 2024/25 season; see Table~\ref{tab:mod_dc_2425}) for the UCL and $\hat{\rho} = 0.064$ for the UEL based on the 2024/25 and 2025/26 seasons (see Table~\ref{tab:mod_dc_combined}).

\begin{figure}[htp!]
    \centering
    \subfigure[Round of 16]{\includegraphics[width=0.48\textwidth]{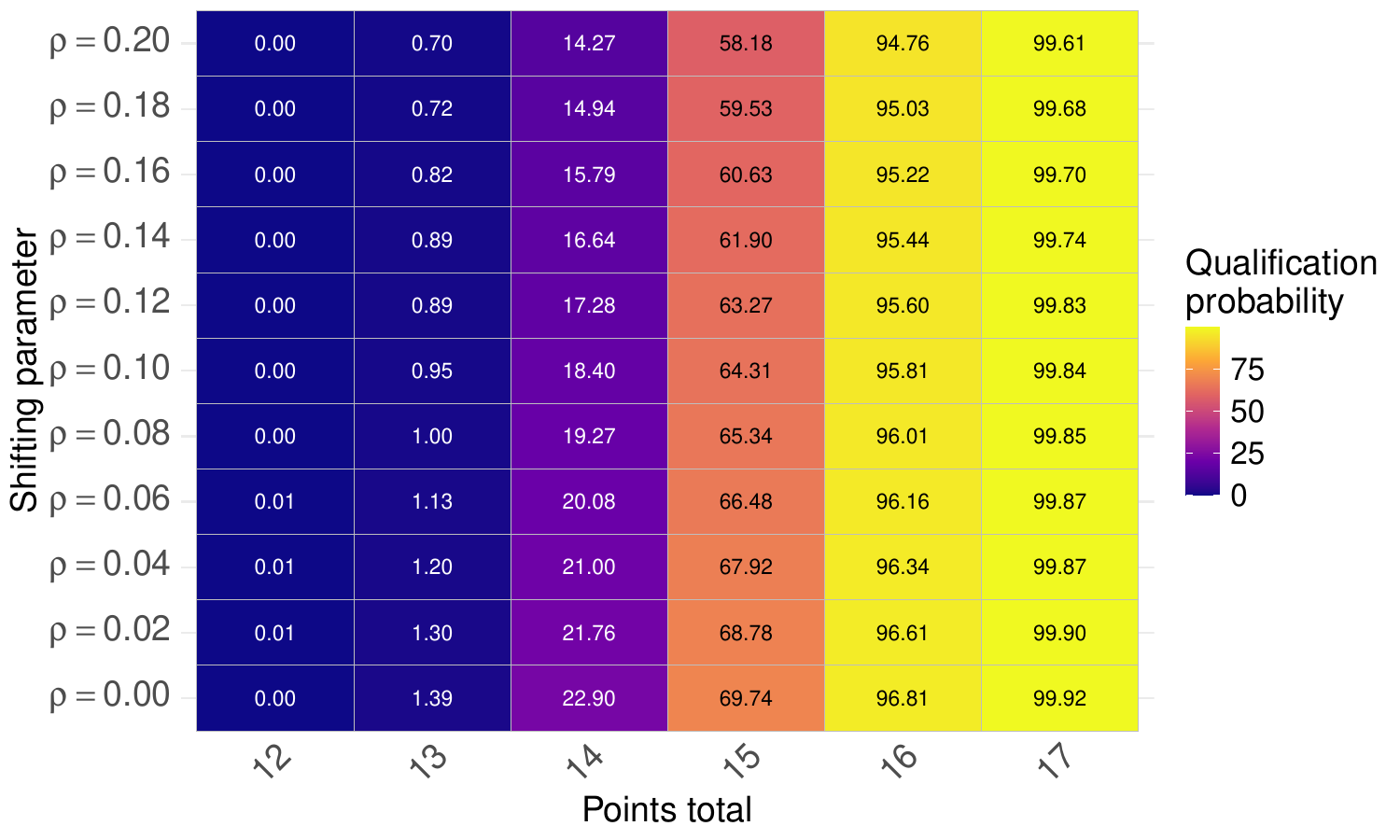}}
    \subfigure[Play-off]
    {\includegraphics[width=0.48\textwidth]{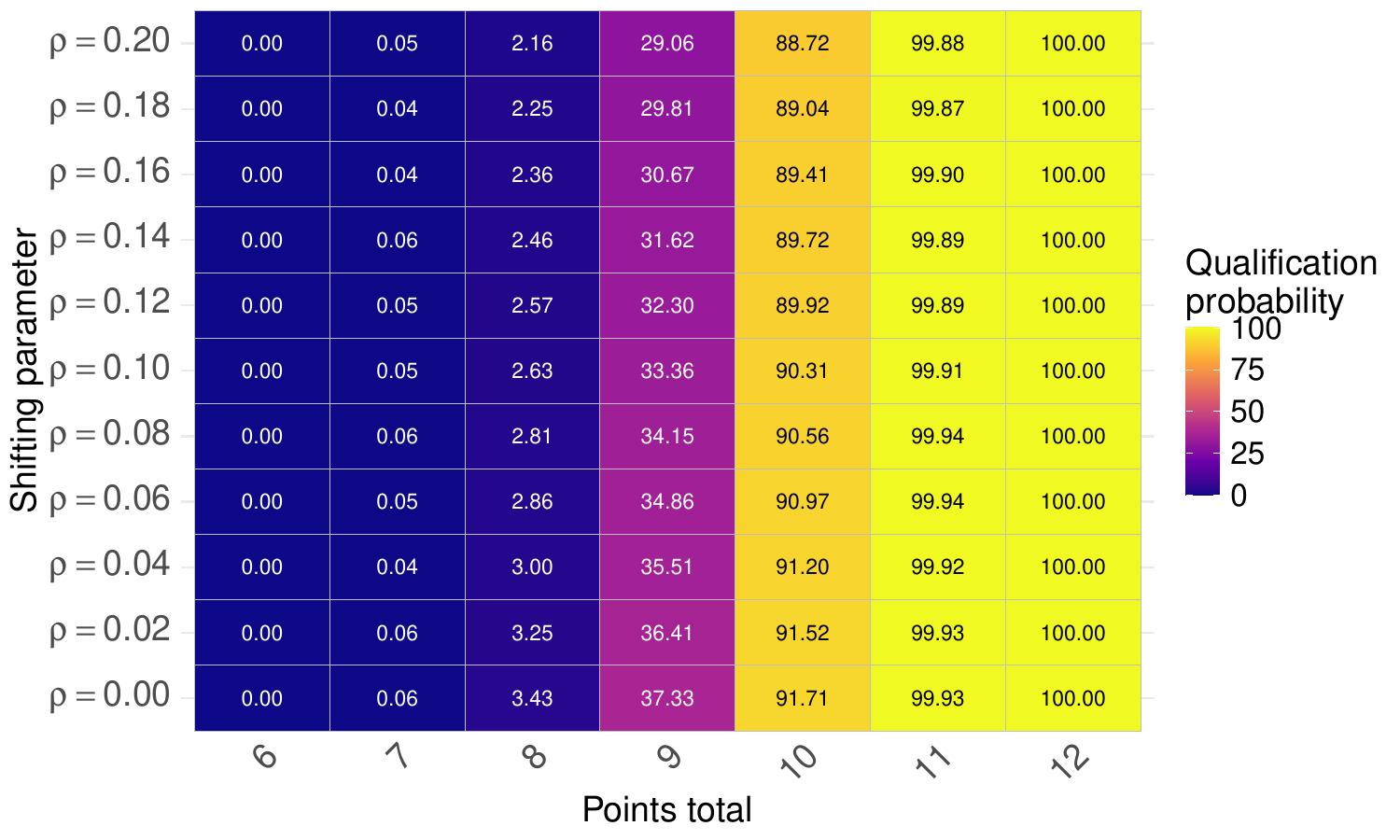}}
    \caption{Qualification probabilities for the Round of 16 (left figure) and play-off phase (right figure) in the UCL depending on the parameter value $\rho$ for randomly generated sets of participating clubs and match schedules.}
    \label{fig:rho_cl}
\end{figure}

To illustrate how this affects qualification thresholds, Figure~\ref{fig:rho_cl} shows the relationship between the value of the parameter shifting probability mass from draws to wins $\rho$ and the qualification probabilities for specific point totals in the UCL, both for progression to the Round of 16~(a) and play-offs~(b). While only minor differences arise when qualification probabilities are already close to 100\% (16 points and above) or 0\% (13 points or less), substantial differences emerge for 14 (15) points. For $\rho = 0$, we obtain a qualification probability of approximately 23\% (70\%); however, the likelihood decreases to around 14\% (58\%) for $\rho = 0.2$. A similar pattern is observed for the play-off round, where the qualification probability at 9 points reduces from 37\% to 29\%. The corresponding results for the UEL are provided in Figure~\ref{fig:rho_el} in Appendix~E. These findings demonstrate that the probability mass on draws can influence qualification probabilities at critical point totals by up to 12 percentage points, underscoring the importance of employing a prediction model capable of integrating the frequency of draws in the new UEFA competition formats to compute thresholds for qualification accurately.

\section{Discussion and managerial implications}
\label{sec:conclusion}
In the 2024/25 season, UEFA introduced an incomplete round-robin format for the Champions League (UCL) and Europa League (UEL), with the aim of increasing competitiveness and fan engagement. Given the relevance of progressing to the next round for strategic planning of clubs, this creates a need for scientific guidance for clubs and managers on qualification thresholds. This paper seeks to develop a model that improves upon existing prediction approaches, which lack validation from the actual outcomes of the 2024/25 UCL season. For this purpose, we consider the difference in competitive balance between the UCL and UEL and explicitly account for the increased incentive to play for a win rather than settle for a draw under the new format by shifting probability mass from draws to wins using the bivariate Dixon--Coles model when predicting qualification thresholds. 

Our results demonstrate that this approach allows us to predict outcomes more accurately. For instance, for a point total of 15, we obtain a qualification probability for the Round of 16 of about 20\% in the 2024/25 UCL season compared to more than 70\% predicted by Opta. Indeed, all teams finishing with 15 points in the UCL missed direct qualification for the Round of 16 and had to compete in the play-offs. While the approach used in this study offers several advantages -- most notable its ability to adjust for unexpected draw rates and account for differences in team strength using Elo ratings -- the first UCL season under the new format may present an extreme case in this regard, requiring caution when interpreting results. To address this, we have not only extended our analysis to an out-of-sample simulation with data from the 2025/26 season, but also presented a simulation based on randomly generated clubs and match schedules, as well as conducted a sensitivity analysis with alternative scenarios that result in different thresholds. These thresholds provide valuable guidelines for teams in planning and refining their strategies. Results from an analysis using market values instead of Elo ratings (see Appendix~D.2) confirm the suitability of the Dixon--Coles model for predicting qualification thresholds in UEFA club competitions. From a managerial perspective, these results enable clubs participating in the UCL or UEL league phase to make better-informed decisions during the winter transfer period regarding their qualification prospects for the play-offs and the Round of 16.

Still, some limitations remain. First, the simulation model does not account for association-specific restrictions in the draw procedure, i.e.\ it does not prevent clubs from being paired with teams from the same association in the league phase. Second, the simulation of matches does not incorporate the schedule of the generated fixtures, as the model relies on static Elo ratings fixed at the beginning of the season. Consequently, dynamic factors such as transfers and roster changes, team motivation, or short-term variations in form, as well as incentives during the final matchdays, are not captured, which may lead to either over- or underestimation of win probabilities. Moreover, the study relies on the two seasons played under the new tournament design. Therefore, the sample size is limited. In addition, it requires simplifying assumptions for the simulation study, such that Elo ratings are uniformly distributed within each seeding pot. 

Despite these limitations, our approach provides valuable guidance for teams and managers. Predicting qualification thresholds can support decision-making regarding transfers, squad rotation, or tactical approaches during the league phase under the uncertainty of progression. Looking ahead, several avenues for further research emerge. Methodologically, the Dixon--Coles model is only one of several approaches for redistributing probabilities across scorelines. With the accumulation of additional seasons, future research could apply more advanced models, such as those proposed by \citet{michels2025extending}. These models not only allow probability mass to be shifted from 0:0 and 1:1 to 0:1 and 1:0, but also enable adjustments for scorelines with more than one goal per team. Such extensions can potentially capture the underlying data-generating process even more accurately, thereby enhancing predictive performance and the robustness of estimated qualification thresholds. In addition, non-linear effects, interaction terms, or more flexible functional forms could be incorporated to better capture the relationship between team strength and scoring intensity.

\vspace{1cm}
\noindent
\textbf{Acknowledgements}: We thank Laszlo Csato for his thoughtful comments and feedback on a previous version of this manuscript.

\bibliographystyle{apalike} 
\bibliography{library}

\newpage
\label{sec:appendix}
\newpage
\section*{Appendix}

\appendix
\begin{appendices}

\renewcommand{\thetable}{A\arabic{table}}
\renewcommand{\thefigure}{A\arabic{figure}}
\setcounter{table}{0}
\setcounter{figure}{0}

\section{UEFA Champions League 2025/26: Final standings}
\label{sec:app_UCL}

\begin{table}[htp!]
    \centering
    \scalebox{0.7}{
    \begin{tabular}{cc|c|c|ccc|ccc|c}
         & & & & & & & Goals & Goals & Goal & \\
         & Place & Team & Matches & Wins & Draws & Losses & scored & conceded & difference & Points \\
         \hline
         \multirow{5}{*}{\rotatebox{90}{\parbox{2.5cm}{\centering Round of 16}}}
         & 1 & Arsenal FC & 8 & 8 & 0 & 0 & 23 & 4 & 19 & 24 \\
         & 2 & Bayern Munich & 8 & 7 & 0 & 1 & 22 & 8 & 14 & 21 \\
         & $\cdots$ & $\cdots$ & $\cdots$ & & $\cdots$ & & & $\cdots$ & & $\cdots$ \\
         & 7 & Sporting CP & 8 & 5 & 1 & 2 & 17 & 11 & 6 & 16 \\
         & 8 & Manchester City & 8 & 5 & 1 & 2 & 15 & 9 & 6 & 16 \\
         \hline
         \multirow{5}{*}{\rotatebox{90}{\parbox{2.5cm}{\centering Play-offs}}}
         & 9 & Real Madrid & 8 & 5 & 0 & 3 & 21 & 12 & 9 & 15 \\
         & 10 & Inter Milan & 8 & 5 & 0 & 3 & 15 & 7 & 8 & 15 \\
         & $\cdots$ & $\cdots$ & $\cdots$ & & $\cdots$ & & & $\cdots$ & & $\cdots$ \\
         & 23 & FK Bodø/Glimt & 8 & 2 & 3 & 3 & 14 & 15 & --1 & 9 \\
         & 24 & SL Benfica & 8 & 3 & 0 & 5 & 10 & 12 & --2 & 9 \\
         \hline
         \multirow{5}{*}{\rotatebox{90}{\parbox{2.5cm}{\centering Elimination}}}
         & 25 & Olympique Marseille & 8 & 3 & 0 & 5 & 11 & 14 & --3 & 9 \\
         & 26 & Pafos FC & 8 & 2 & 3 & 3 & 8 & 11 & --3 & 9 \\
         & $\cdots$ & $\cdots$ & $\cdots$ & & $\cdots$ & & & $\cdots$ & & $\cdots$ \\
         & 35 & FC Villarreal & 8 & 0 & 1 & 7 & 5 & 18 & --13 & 1 \\
         & 36 & Qairat Almaty & 8 & 0 & 1 & 7 & 7 & 22 & --15 & 1 \\
    \end{tabular}}
    \caption{Final standings after the UCL league phase 2025/26.}
    \label{tab:standings_CL_2526}
\end{table}

\newpage
\section{UEFA Europa League 2024/25 and 2025/26:\\ Final standings}
\label{sec:app_UEL}

\renewcommand{\thetable}{B\arabic{table}}
\renewcommand{\thefigure}{B\arabic{figure}}
\setcounter{table}{0}
\setcounter{figure}{0}

\begin{table}[htp!]
    \centering
    \scalebox{0.75}{
    \begin{tabular}{ccc|c|c|ccc|ccc|c}
         & & & & & & & & Goals & Goals & Goal & \\
         & & Place & Team & Matches & Wins & Draws & Losses & scored & conceded & difference & Points \\
         \hline
          \multirow{15}{*}{\rotatebox{90}{\parbox{2.5cm}{\centering 2024/25}}}
         & \multirow{5}{*}{\rotatebox{90}{\parbox{2.5cm}{\centering Round of 16}}}
         &  1 & Lazio & 8 & 6 & 1 & 1 & 17 & 5 & 12 & 19 \\
         & & 2 & Athletic Club & 8 & 6 & 1 & 1 & 15 & 7 & 8 & 19 \\
         & & $\cdots$ & $\cdots$ & $\cdots$ & & $\cdots$ & & & $\cdots$ & & $\cdots$ \\
         & & 7 & Olympiacos & 8 & 4 & 3 & 1 & 9 & 3 & 6 & 15 \\
         & & 8 & Rangers & 8 & 4 & 2 & 2 & 16 & 10 & 6 & 14 \\
         \hline
         & \multirow{5}{*}{\rotatebox{90}{\parbox{2.5cm}{\centering Play-offs}}}         
         & 9 & FK Bodø/Glimt & 8 & 4 & 2 & 2 & 14 & 11 & 3 & 14 \\
         & & 10 & Anderlecht & 8 & 4 & 2 & 2 & 14 & 12 & 2 & 14 \\
         & & $\cdots$ & $\cdots$ & $\cdots$ & & $\cdots$ & & & $\cdots$ & & $\cdots$ \\
         & & 23 & Twente & 8 & 2 & 4 & 2 & 8 & 9 & --1 & 10 \\
         & & 24 & Fenerbahce & 8 & 2 & 4 & 2 & 9 & 11 & --2 & 10 \\
         \hline
         & \multirow{5}{*}{\rotatebox{90}{\parbox{2.5cm}{\centering Elimination}}}
         & 25 & Braga & 8 & 3 & 1 & 4 & 9 & 12 & --3 & 10 \\
         & & 26 & Elfsborg & 8 & 3 & 1 & 4 & 9 & 14 & --5 & 10 \\
         & & $\cdots$ & $\cdots$ & $\cdots$ & & $\cdots$ & & & $\cdots$ & & $\cdots$ \\
         & & 35 & Nice & 8 & 0 & 3 & 5 & 7 & 16 & --9 & 3 \\
         & & 36 & Qarabag & 8 & 1 & 0 & 7 & 6 & 20 & --14 & 3 \\

        \hline
        \hline

         \multirow{15}{*}{\rotatebox{90}{\parbox{2.5cm}{\centering 2025/26}}}
        & \multirow{5}{*}{\rotatebox{90}{\parbox{2.5cm}{\centering Round of 16}}}
        & 1 & Olympique Lyon & 8 & 7 & 0 & 1 & 18 & 5 & 13 & 21 \\
        & & 2 & Aston Villa & 8 & 7 & 0 & 1 & 14 & 6 & 8 & 21 \\
        & & $\cdots$ & $\cdots$ & $\cdots$ &  & $\cdots$ &  &  & $\cdots$ &  & $\cdots$ \\
        & & 7 & SC Freiburg & 8 & 5 & 2 & 1 & 10 & 4 & 6 & 17 \\
        & & 8 & AS Rom & 8 & 5 & 1 & 2 & 13 & 6 & 7 & 16 \\
        \cline{2-12}
        & \multirow{5}{*}{\rotatebox{90}{\parbox{2.5cm}{\centering Play-offs}}}
        & 9 & KRC Genk & 8 & 5 & 1 & 2 & 11 & 7 & 4 & 16 \\
        & & 10 & FC Bologna & 8 & 4 & 3 & 1 & 14 & 7 & 7 & 15 \\
        & & $\cdots$ & $\cdots$ & $\cdots$ &  & $\cdots$ &  &  & $\cdots$ &  & $\cdots$ \\
        & & 23 & GNK Dinamo Zagreb & 8 & 3 & 1 & 4 & 12 & 16 & -4 & 10 \\
        & & 24 & Brann Bergen & 8 & 2 & 3 & 3 & 9 & 11 & -2 & 9 \\
        \cline{2-12}
        & \multirow{5}{*}{\rotatebox{90}{\parbox{2.5cm}{\centering Elimination}}}
        & 25 & BSC Young Boys & 8 & 3 & 0 & 5 & 10 & 16 & -6 & 9 \\
        & & 26 & Sturm Graz & 8 & 2 & 1 & 5 & 5 & 11 & -6 & 7 \\
        & & $\cdots$ & $\cdots$ & $\cdots$ &  & $\cdots$ &  &  & $\cdots$ &  & $\cdots$ \\
        & & 35 & Malmö FF & 8 & 0 & 1 & 7 & 4 & 15 & -11 & 1 \\
        & & 36 & Maccabi Tel Aviv & 8 & 0 & 1 & 7 & 2 & 22 & -20 & 1 \\
    \end{tabular}}
    \caption{Final standings after the UEL league phase 2024/25 and 2025/26.}
    \label{tab:standings_EL}
\end{table}

\newpage
\section{Elo rating of UCL clubs}

\renewcommand{\thetable}{C\arabic{table}}
\renewcommand{\thefigure}{C\arabic{figure}}
\setcounter{table}{0}
\setcounter{figure}{0}

\begin{table}[htp!]
\centering
\scalebox{0.7}{
\begin{tabular}{lcc}
\hline
Club & Elo rating 01.09.2024 & Elo rating 01.09.2025 \\
\hline
AC Milan & 1818 & -- \\
Ajax & -- & 1662 \\
Arsenal FC & 1950 & 1996 \\
Aston Villa & 1778 & -- \\
Atalanta BC & 1866 & 1830 \\
Athletic Club & -- & 1805 \\
Atletico Madrid & 1838 & 1833 \\
Bayer Leverkusen & 1918 & 1832 \\
Bayern Munich & 1905 & 1930 \\
Borussia Dortmund & 1870 & 1820 \\
BSC Young Boys & 1553 & -- \\
Celtic & 1644 & -- \\
Chelsea & -- & 1915 \\
Club Brugge KV & 1703 & 1754 \\
Copenhagen & -- & 1663 \\
Crvena Zvezda (Red Star Belgrade) & 1566 & -- \\
Eintracht Frankfurt & -- & 1764 \\
FC Barcelona & 1898 & 1948 \\
FC Bologna & 1769 & -- \\
FC Villarreal & -- & 1794 \\
Feyenoord & 1748 & -- \\
FK Bodø/Glimt & -- & 1655 \\
Galatasaray & -- & 1711 \\
Girona & 1792 & -- \\
GNK Dinamo Zagreb & 1586 & -- \\
Inter Milan & 1966 & 1923 \\
Juventus & 1836 & 1816 \\
LOSC Lille & 1779 & -- \\
Liverpool FC & 1908 & 2010 \\
Manchester City & 2060 & 1939 \\
Monaco & 1773 & 1762 \\
Napoli & -- & 1846 \\
Newcastle United & -- & 1862 \\
Olympiacos & -- & 1691 \\
Olympique Marseille & -- & 1743 \\
Pafos FC & -- & 1525 \\
Paris Saint-Germain & 1887 & 1979 \\
PSV Eindhoven & 1795 & 1786 \\
Qairat Almaty & -- & 1302 \\
Qarabag & -- & 1524 \\
RB Leipzig & 1861 & -- \\
Red Bull Salzburg & 1686 & -- \\
Real Madrid & 1985 & 1944 \\
Shakhtar Donetsk & 1572 & -- \\
SL Benfica & 1759 & 1821 \\
Slavia Prague & -- & 1680 \\
Slovan Bratislava & 1446 & -- \\
Sporting CP & 1835 & 1793 \\
Sparta Prague & 1728 & -- \\
Stade Brestois & 1685 & -- \\
Sturm Graz & 1606 & -- \\
Tottenham Hotspur & -- & 1794 \\
Union Saint-Gilloise & -- & 1733 \\
VfB Stuttgart & 1791 & -- \\
\hline
\end{tabular}
}
\caption{Elo ratings as of 01.09.2024 and 01.09.2025 for clubs participating in the 2024/25 and 2025/26 UCL season, respectively, obtained from~\url{www.clubelo.com}.}
\label{tab:elo}
\end{table}

\renewcommand{\thetable}{C\arabic{table}}
\renewcommand{\thefigure}{C\arabic{figure}}
\setcounter{table}{0}
\setcounter{figure}{0}

\newpage
\section{Model comparison}
\label{sec:app_comparison}

\renewcommand{\thetable}{D\arabic{table}}
\renewcommand{\thefigure}{D\arabic{figure}}
\setcounter{table}{0}
\setcounter{figure}{0}

\subsection{Model frameworks}

To assess whether the Dixon--Coles model provides the most suitable structural specification for predicting match outcomes in the new UEFA club competition format, we compare its forecasting performance with the independent Poisson and the well-known bivariate Poisson model \citep{karlis2005bivariate}. 

Models are evaluated using strictly proper scoring rules \citep{gneiting2007strictly} to assess their probabilistic forecasting performance. In particular, we consider the Brier Score, the Ranked Probability Score (RPS; \citealp{constantinou2012solving}), and the Logarithmic Score (Log Loss). The Brier Score measures the mean squared deviation between predicted probabilities and observed outcomes in a multi-class setting. The RPS extends this concept to ordered categorical outcomes and evaluates cumulative probability discrepancies, making it particularly suitable for football results (home win, draw, away win). The Logarithmic Score evaluates the log-likelihood assigned to the realised outcome and is closely related to maximum likelihood estimation. All three scoring rules are strictly proper, meaning that they are minimised in expectation by the true data-generating distribution \citep{gneiting2007strictly}. Lower values of these metrics indicate superior probabilistic calibration and sharpness. The comparison is conducted out of sample, using the 2024/25 season as training data and the 2025/26 season as test data. We do not train on the latter season and test on the former one to prevent data leakage.

\begin{table}[ht]
\centering
\begin{tabular}{|l|ccc|ccc|}
  \hline
  & \multicolumn{3}{c|}{UCL} & \multicolumn{3}{c|}{UEL} \\ 
  \hline
  Metric & Ind.\ Pois. & D\&C & Biv.\ Pois. & Ind.\ Pois. & D\&C & Biv.\ Pois. \\ 
  \hline
Brier Score 
& 0.5268 & \textbf{0.5262} & 0.5300 
& 0.5533 & \textbf{0.5531} & 0.5574 \\ 

RPS 
& 0.3868 & \textbf{0.3865} & 0.3883 
& 0.4089 & 0.4088 & \textbf{0.4085} \\ 

Log Score 
& 0.8988 & \textbf{0.8966} & 0.9074 
& 0.9415 & \textbf{0.9412} & 0.9505 \\ 
\hline
\end{tabular}
\caption{Out-of-sample predictive performance of the independent Poisson (Ind.\ Pois.), bivariate Poisson (Biv.\ Pois.), and Dixon--Coles (D\&C) model for the UCL and UEL, evaluated by the Brier Score, Ranked Probability Score (RPS), and Logarithmic Score.}
\label{tab:comp}
\end{table}
Table~\ref{tab:comp} reports the out-of-sample predictive performance of the competing models for both competitions, the UCL and the UEL. For the UCL, the Dixon--Coles specification achieves the best performance across all three scoring rules. Compared to the standard, independent Poisson model, the improvements are consistent across all evaluation metrics. The bivariate Poisson model performs worse, suggesting that the dependence structure introduced by this specification does not lead to improved predictive accuracy in this setting.

For the UEL sample, the results are similar. The Dixon--Coles model achieves the lowest Brier Score (0.5531) and Logarithmic Score (0.9412), with only marginal differences compared to the independent Poisson model (0.5533 and 0.9415, respectively). Regarding the RPS, the bivariate Poisson model performs marginally best (0.4085), although the differences between the three models are negligible.

Overall, the results indicate that the Dixon--Coles model provides the most robust predictive performance among the considered specifications, particularly in the UCL, consistent across different scoring rules. This suggests that accounting for low-score dependence, as introduced by \citet{dixon1997modelling}, is important when forecasting unseen matches in the new UEFA club competition tournament format.

\subsection{Covariate modelling}

\begin{figure}[htp!]
    \centering
    \subfigure[UCL]{\includegraphics[width=0.48\textwidth]{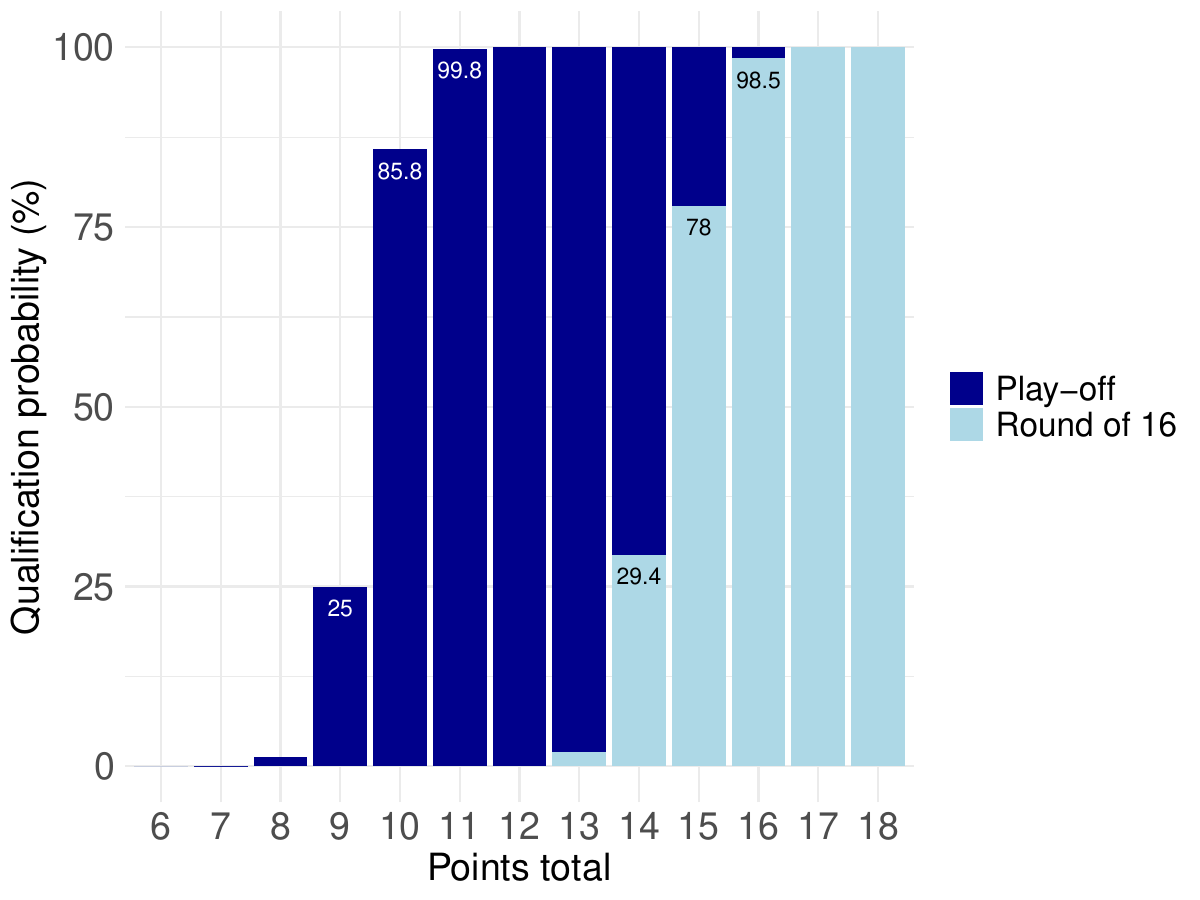}}
    \subfigure[UEL]
    {\includegraphics[width=0.48\textwidth]{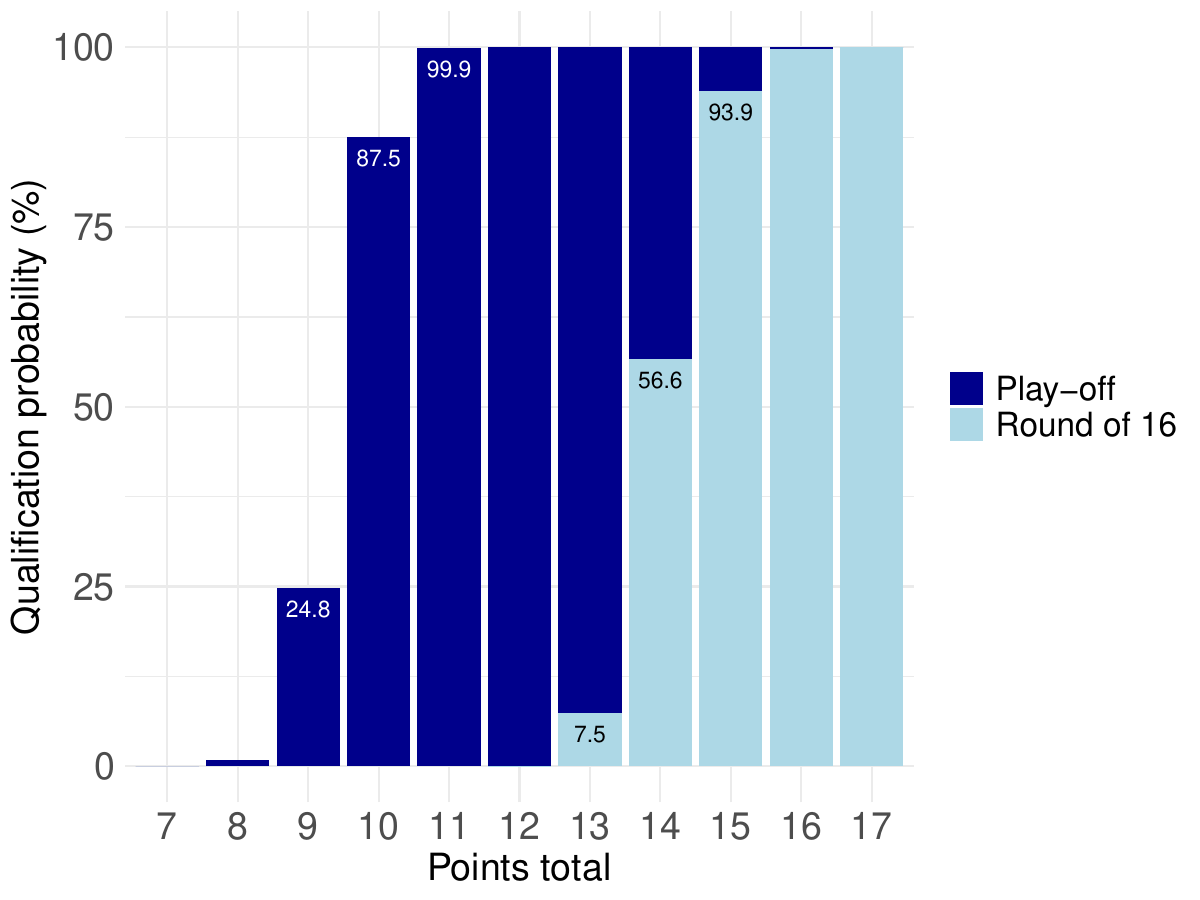}}
    \caption{Probability to progress to the Round of 16 and play-offs, respectively, in the 2025/26 season (out-of-sample simulation), depending on the number of points achieved. The prediction is based on the win probabilities estimated from the 2024/25 season schedule with the Dixon--Coles model using market values (Table~\ref{tab:mod_dc_cl_mv}).}
    \label{fig:clel_dc_2526_mv}
\end{figure}

To ensure that our conclusions are not driven by the specific choice of team-strength proxy, we again predict qualification thresholds in an out-of-sample analysis using data from the 2024/25 season and evaluate the 2025/26 season for model comparison (similar to Section~\ref{sec:predict_2526}). Instead of relying on Elo ratings, we now use market values (MVs) as an alternative explanatory variable. While the suitability of Elo ratings for European club tournaments has been demonstrated \citep{csato2024club}, we do not claim that they are a perfect proxy for match outcomes. Similarly, MVs are only an indirect measure of team quality. However, the objective of this robustness check is not to identify the optimal strength indicator, but rather to examine whether the model comparison remains stable across different covariate specifications. To account for the multiplicative effect of market values, we slightly adjust the model formulation from Section~\ref{sec:predictor}. Specifically, we estimate the Dixon--Coles model with the following properties:
\begin{align*}
\lambda &= \exp\left(\beta_0 + \beta_1 \cdot \left( \log\left(\text{MV}_{\text{Home}} \right) - \log\left(\text{MV}_{\text{Away}} \right) \right) + \beta_2 \right),\\
\mu &= \exp\left(\beta_0 - \beta_1 \cdot \left( \log\left(\text{MV}_{\text{Home}} \right) - \log\left(\text{MV}_{\text{Away}} \right) \right) \right).
\end{align*}
Table~\ref{tab:mod_dc_cl_mv} provides estimated coefficients for the model fitted to the 2024/25 UCL season (left columns) and to both seasons combined (right columns). The parameter shifting probability mass from draws to 1:0 and 0:1 is estimated as 0.109 (0.0001) for the 2024/25 season in the UCL (UEL) and as 0.041 (0.062) for both seasons combined. This is fairly close to the values obtained with Elo ratings of 0.102 (0.005) for the 2024/25 season (see Table~\ref{tab:mod_dc_2425} in Section~\ref{sec:predict_2526}) and 0.036 (0.064) for both seasons (see Table~\ref{tab:mod_dc_combined} in Section~\ref{sec:sim}). Unsurprisingly, for most point totals, predicted qualification thresholds for the 2025/26 reported in Figure~\ref{fig:clel_dc_2526_mv} are similar to those obtained when using Elo ratings (see Figure~\ref{fig:clel_dc_2526} in Section~\ref{sec:predict_2526}). Specifically, deviations in qualification probabilities are smaller than six percentage points in the UCL, with one exception: When using Elo ratings, we predict a probability of 41.4\% to advance to the play-offs with nine points, whereas this probability is predicted at 25.0\% when using MVs. Indeed, in the 2025/26 season, two out of five teams (40\%) with nine points reached the play-offs. 

\begin{table}[!htbp] \centering 
  \caption{Estimated coefficients for the Dixon--Coles models fitted to the 2024/25 UCL and UEL season (left columns) and both seasons combined (right columns) with market values instead of Elo ratings.} 
  \label{tab:mod_dc_cl_mv} 
  \scalebox{0.75}{
\begin{tabular}{@{\extracolsep{5pt}}lcc|cc} 
\\[-1.8ex]\hline 
\hline \\[-1.8ex] 
 & \multicolumn{4}{c}{\textit{Dependent variable: Goals}} \\ 
\cline{2-5}
 & \multicolumn{2}{c|}{\textit{2024/25}} 
   & \multicolumn{2}{c}{\textit{Both seasons}} \\ 
 & UCL & UEL & UCL & UEL \\ 
\hline 
 MVDiff & 0.295$^{***}$ & 0.149$^{***}$ & 0.234$^{***}$ & 0.184$^{***}$ \\ 
  & (0.033) & (0.033) & (0.021) & (0.026) \\ 
  & & & & \\ 
 Home & 0.295$^{***}$ & 0.316$^{***}$ & 0.290$^{***}$ & 0.293$^{***}$ \\ 
  & (0.094) & (0.100) & (0.065) & (0.072) \\ 
  & & & & \\ 
 $\rho$ & 0.109 & 0.0001 & 0.041 & 0.062 \\ 
  & (0.096) & (0.112) & (0.070) & (0.078) \\ 
  & & & & \\ 
 Constant & 0.240$^{***}$ & 0.152$^{**}$ & 0.284$^{***}$ & 0.132$^{**}$ \\ 
  & (0.073) & (0.077) & (0.051) & (0.055) \\ 
  & & & & \\ 
\hline 
Observations & 144 & 144 & 288 & 288 \\ 
\hline
\hline \\[-1.8ex] 
\multicolumn{5}{l}{\textit{Note:} $^{*}p<0.1; \, ^{**}p<0.05; \, ^{***}p<0.01$} \\ 
\end{tabular}}
\end{table}

\newpage

\section{Qualification probabilities depending on $\rho$ for the UEL}

\renewcommand{\thetable}{E\arabic{table}}
\renewcommand{\thefigure}{E\arabic{figure}}
\setcounter{table}{0}
\setcounter{figure}{0}

\begin{figure}[htp!]
    \centering
    \subfigure[Round of 16]{\includegraphics[width=0.48\textwidth]{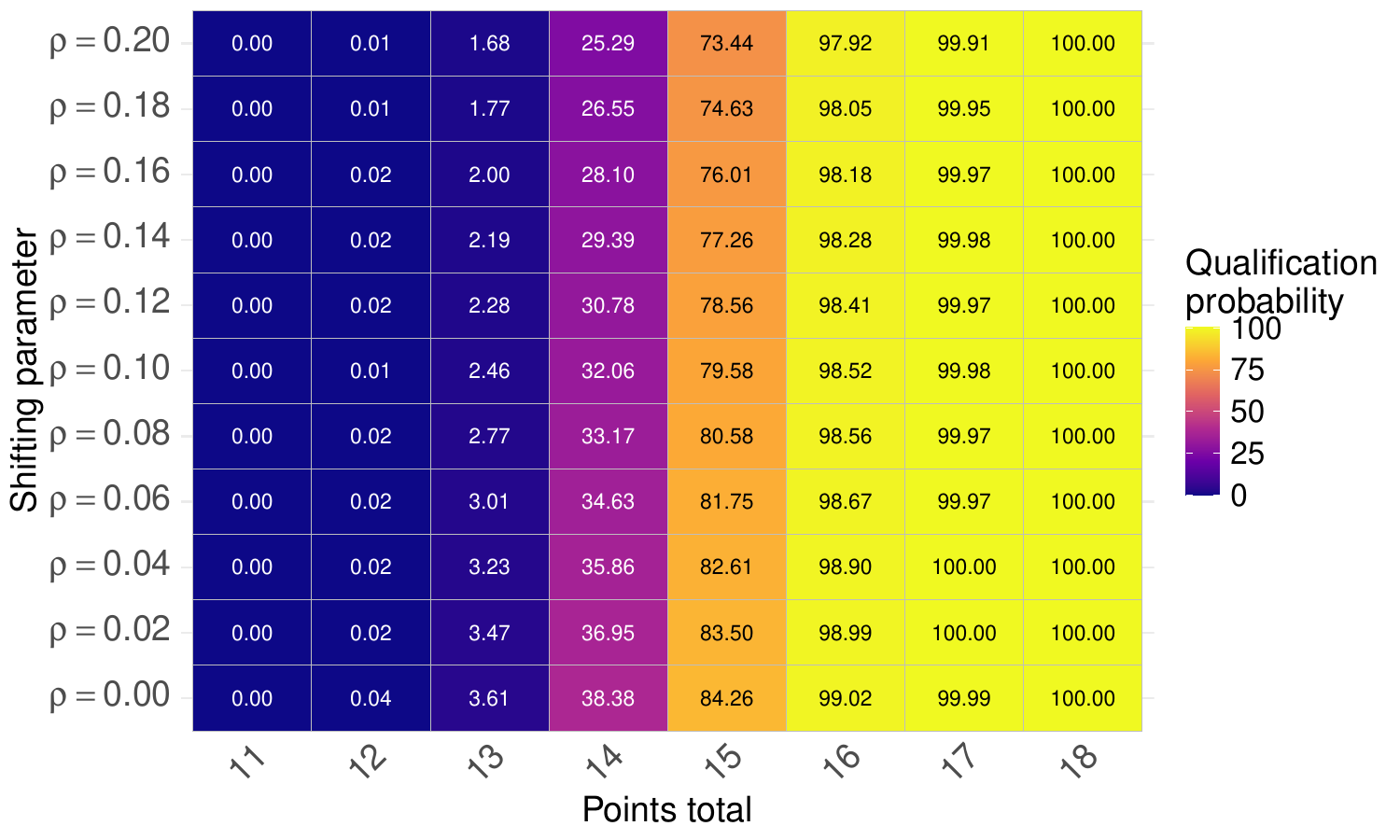}}
    \subfigure[Play-off]
    {\includegraphics[width=0.48\textwidth]{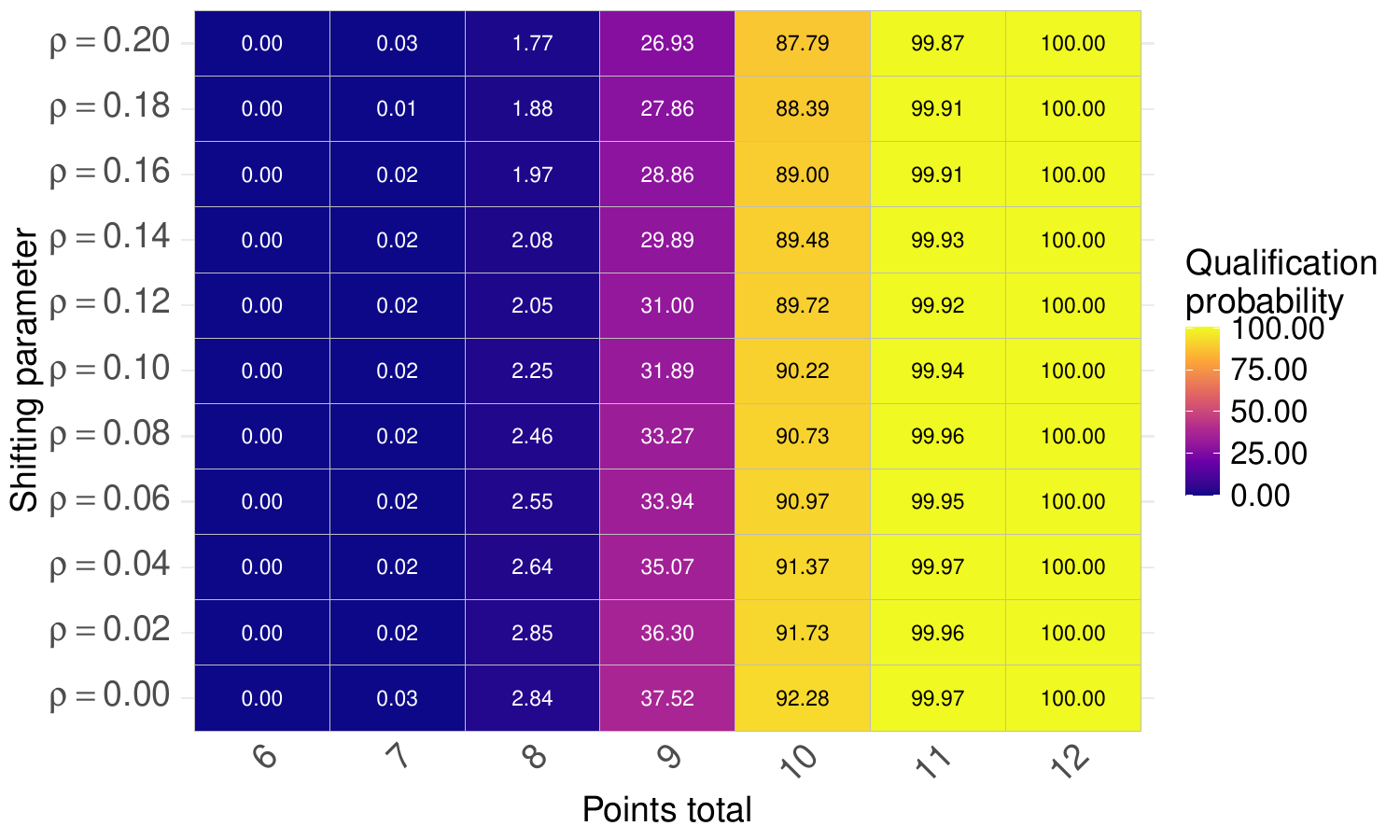}}
    \caption{Qualification probabilities for the Round of 16 (left figure) and play-off phase (right figure) in the UEL with randomly drawn match schedules and different values for the parameter $\rho$.}
    \label{fig:rho_el}
\end{figure}

\end{appendices}

\end{spacing}
\end{document}